\documentclass[11pt,preprint,aps,showpacs]{revtex4-1}
\usepackage{mathrsfs}
\usepackage{amsmath}
\usepackage{amssymb}
\usepackage{epsfig}
\usepackage{graphicx}
\usepackage{booktabs}
\usepackage{array}
\usepackage{multirow}
\usepackage{slashed}
\usepackage{color}
\usepackage{makecell}
\usepackage{setspace}
\usepackage{array} 
\usepackage{caption}
\usepackage{subcaption}
\usepackage{bm}
\usepackage{mathtools}
\usepackage[colorlinks=true,linkcolor=red,citecolor=blue]{hyperref}
\allowdisplaybreaks[4]
\newcommand{\wm}{\phantom{-}}
\begin{document}
	\newcommand{\non}{\nonumber\\ }
	
	\title{The $\Lambda_{b} \to \Lambda$ transition form factors in perturbative QCD approach}
	\author{
		Lei Yang$^1$ \footnote{leiyang@htu.edu.cn},
		Jia-Jie Han$^2$ \footnote{hanjj@lzu.edu.cn, corresponding author},
		Qin Chang$^{1,3}$ \footnote{changqin@htu.edu.cn, corresponding author},
		Fu-Sheng Yu$^2$ \footnote{yufsh@lzu.edu.cn, corresponding author}
	}
	\affiliation
	{\small
		1.~Centre for Theoretical Physics, Henan Normal University, Xinxiang 453007, China\\
		2.~Frontiers Science Center for Rare Isotopes, and School of Nuclear Science and Technology, Lanzhou University, Lanzhou 730000,  China \\
		3.~Center for High Energy Physics, Henan Academy of Sciences, Zhengzhou 455004, China
	}

	\begin{abstract}
		
		We perform a systematic analysis of the $\Lambda_b \to \Lambda$ transition form factors using the perturbative QCD (PQCD) approach, taking into account contributions from higher-twist light-cone distribution amplitudes (LCDAs). Using inputs from lattice QCD, we show that the baryon higher-twist LCDAs give numerically dominant contributions to the form factors. By combining our PQCD results at low-$q^2$ region with lattice QCD predictions at high $q^2$, we carry out $z$-series expansion fits to obtain a unified description of the form factors over the full physical kinematic range. We also provide predictions for physical observables in the rare decay $\Lambda_b \to \Lambda \mu^+ \mu^-$, such as the differential branching fraction, the longitudinal polarization fraction of the dimuon system, and the forward–backward asymmetries. 
		
	\end{abstract}
	\maketitle
	\section{introduction}
	
	In the Standard Model (SM), processes induced by flavor-changing neutral currents (FCNCs), such as the $b \to s$ transition, are strongly suppressed by the Glashow–Iliopoulos–Maiani (GIM) mechanism~\cite{Glashow:1970gm} since they proceed only through loop diagrams. These FCNC decays thus serve as sensitive probes of the SM, as potential new physics contributions could alter their observable signatures.
    Several deviations between theoretical predictions and experimental measurements have been reported by the LHCb Collaboration in $b \to s l^{+}l^{-}$ transitions, including the decays $B^{+} \to K^{(*)+} \mu^{+} \mu^{-}$, $B^{0} \to K^{0} \mu^{+} \mu^{-}$~\cite{LHCb:2014cxe}, $B^{0} \to K^{*0} \mu^{+} \mu^{-}$~\cite{LHCb:2016ykl}, $B^{0}_{s} \to \phi \mu^{+} \mu^{-}$~\cite{LHCb:2021zwz}, and the baryonic mode $\Lambda_{b}^{0} \to \Lambda^0 \mu^{+} \mu^{-}$~\cite{LHCb:2015tgy}. Such anomalies motivate detailed theoretical studies of the underlying transition form factors.

	
    The accurate determination of form factors is also essential in predicting CP-violating observables in nonleptonic modes.
    The LHCb collaboration has recently reported the first observation of CP violation in $\Lambda_b\to pK^-\pi^+\pi^-$ decay and reported a large local CP asymmetry exceeding $6.0\sigma$ significance in the region of $m_{p\pi^+\pi^-}<2.7\text{GeV}/c^2$~\cite{LHCb:2025ray}, which significantly advance our understanding of CP violation in baryon decays. The LHCb measurements implie the importance of dynamical studies of $\Lambda_b$ decays by the consistency between the experimental results~\cite{LHCb:2025ray} and the theoretical predictions within the $N\pi$ rescattering mechanism~\cite{Wang:2024oyi}.  
	Dominant decay modes of $\Lambda$ baryon to nucleon and pion provide substantial strong phases that may enhance local CP violation, as the interference between resonances is known to play an important role in generating large local CP violation~\cite{Wang:2024oyi,Yu:2025ekh}. The LHCb Collaboration has also found evidence of CP violation in the three-body decay $\Lambda_b\to \Lambda K^+K^-$~\cite{LHCb:2024yzj}. These experimental breakthroughs motivate our systematic investigation of the $\Lambda_b\to\Lambda$ transition form factors, aiming to establish a robust theoretical foundation for interpreting these intriguing observations in the future works.
    
    Several theoretical approaches have been applied to calculate the $\Lambda_b\to \Lambda$ form factors, including the soft-collinear effective theory~\cite{Feldmann:2011xf}, the lattice QCD~\cite{Detmold:2012vy,Detmold:2016pkz}, and the light-cone sum rules~\cite{Wang:2015ndk}, each offering distinct advantages in describing the form factors.
    Among various theoretical approaches, the perturbative QCD (PQCD) approach, preserving the transverse momentum based on the $k_T$ factorization, has proven effective in studying heavy-to-light hadronic transitions at the fast recoil region~\cite{Kurimoto:2001zj,Li:2012nk}. The PQCD approach is especially valuable as it can calculate nonfactorizable contributions unambiguously, such as the $W$-exchange and penguin-annihilation amplitudes, which are essential for studying CP-violating observables in nonleptonic decays~\cite{Lu:2000em,Cheng:2014rfa}. Thus, the PQCD approach has been successfully applied in predicting the CP violations of $B$ meson decays~\cite{Belle:2004mad,BaBar:2004gyj,Belle:2004nch}, and recently explaining the long-standing puzzle of why the measured CP violations in $\Lambda_b\to p\pi^-,pK^-$ are so small compared to their $B$ meson counterparts~\cite{LHCb:2018fly,LHCb:2024iis,Han:2024kgz,Han:2025tvc}.

	The first application of the PQCD approach to form factor calculations was carried out in~\cite{Li:1992ce,Kundu:1998gv}, where the proton electromagnetic form factors were evaluated with Sudakov suppression. Subsequent works~\cite{Shih:1998pb,Lu:2009cm} extended this framework to the $\Lambda_b \to p$ transition, though the predicted form factors were notably smaller than those from nonperturbative methods. More recently, the inclusion of high-twist baryon LCDAs in~\cite{Han:2022srw} has helped reconcile this discrepancy, reproducing $\Lambda_b \to p$ form factors of reasonable magnitude and thereby reinforcing the validity of the PQCD approach.
	
	The form factors of the $V-A$ current for the $\Lambda_b \to \Lambda$ transition were studied in~\cite{Rui:2022sdc}, where different models for the $\Lambda$ baryon LCDAs were examined. The model derived from lattice QCD~\cite{Bali:2015ykx,RQCD:2019hps} only account for leading-twist LCDAs, leading to excessively small form factors. Another choice for the $\Lambda$ baryon LCDAs is from QCD sum rule (QCDSR)~\cite{Liu:2014uha}, which includes high-power corrections up to twist-6 and next-to-leading-order conformal spin expansion contributions. However, the resulting form factors are anomalously large and violate heavy-quark symmetry~\cite{Rui:2022sdc}, suggesting that the $\Lambda$ baryon LCDAs obtained in~\cite{Liu:2014uha} by QCDSR may contain problematic assumptions.
	
	In this work, we systematically investigate all ten form factors describing the $\Lambda_b\to \Lambda$ transition induced by the $b\to s$ vector, axial-vector, and tensor currents within the PQCD approach. Our calculations successfully reproduce $\Lambda_b\to \Lambda$ form factors with reasonable magnitude, consistent with those obtained from other theoretical methods. We further show that high-twist baryon LCDAs provide the dominant contribution to baryonic transition form factors, which aligns with the conclusion drawn in~\cite{Han:2022srw}.
	
	The remainder of this paper is organized as follows. In Sec.~\ref{definitionAndPQCD}, we introduce the definitions of the $\Lambda_b \to \Lambda$ transition form factors and presents the PQCD framework for baryonic transition form factors. The LCDAs for the $\Lambda_b$ and $\Lambda$ baryons used in this work are listed in Sec.~\ref{sec:LCDAs} in detail. Numerical results for the form factors are presented in Sec.~\ref{sec:FFs}, where we also perform extrapolations using the modified $z$-expansion. Our predictions for the differential branching fraction and angular observables in $\Lambda_b \to \Lambda(\to p\pi^-)\mu^+\mu^-$ decays are given in Sec.~\ref{sec:observables}. Finally, we conclude with a summary in Sec.~\ref{sec:summary}. The Appendix provides the explicit factorization formulas for each relevant diagram.
	
	\section{Definition Of The Form Factors and PQCD framework}\label{definitionAndPQCD}
	The helicity-based definition of the $\Lambda_b\to \Lambda$ form factors are given by~\cite{Feldmann:2011xf}
	\begin{align}
		\langle \Lambda(p',s') | \overline{s} \gamma^\mu b | \Lambda_b(p,s) \rangle=& \overline{u}_\Lambda(p',s') \Big[
		f_0(q^2)\, (m_{\Lambda_b} - m_\Lambda)\frac{q^\mu}{q^2} \notag \\
		&+ f_+(q^2) \frac{m_{\Lambda_b} + m_\Lambda}{s_+}
		\left( p^\mu + p^{\prime \mu} - (m_{\Lambda_b}^2 - m_\Lambda^2)\frac{q^\mu}{q^2} \right) \notag \\
		&+ f_\perp(q^2) \left( \gamma^\mu - \frac{2m_\Lambda}{s_+} p^\mu - \frac{2 m_{\Lambda_b}}{s_+} p^{\prime \mu} \right)
		\Big] u_{\Lambda_b}(p,s)  \\
		\langle \Lambda(p',s') | \overline{s} \gamma^\mu \gamma_5 b | \Lambda_b(p,s) \rangle
		=& -\overline{u}_\Lambda(p',s') \gamma_5 \Big[
		g_0(q^2)\, (m_{\Lambda_b} + m_\Lambda)\frac{q^\mu}{q^2} \notag \\
		&+ g_+(q^2) \frac{m_{\Lambda_b} - m_\Lambda}{s_-}
		\left( p^\mu + p^{\prime \mu} - (m_{\Lambda_b}^2 - m_\Lambda^2)\frac{q^\mu}{q^2} \right) \notag \\
		&+ g_\perp(q^2) \left( \gamma^\mu + \frac{2m_\Lambda}{s_-} p^\mu - \frac{2 m_{\Lambda_b}}{s_-} p^{\prime \mu} \right)
		\Big] u_{\Lambda_b}(p,s)  \\
		\langle \Lambda(p',s') | \overline{s} \, i \sigma^{\mu\nu} q_\nu \, b | \Lambda_b(p,s) \rangle
		=& -\overline{u}_\Lambda(p',s') \Big[
		h_+(q^2) \frac{q^2}{s_+}
		\left( p^\mu + p^{\prime \mu} - (m_{\Lambda_b}^2 - m_\Lambda^2)\frac{q^\mu}{q^2} \right) \notag \\
		&+ h_\perp(q^2)(m_{\Lambda_b} + m_\Lambda)
		\left( \gamma^\mu - \frac{2m_\Lambda}{s_+} p^\mu - \frac{2 m_{\Lambda_b}}{s_+} p^{\prime \mu} \right)
		\Big] u_{\Lambda_b}(p,s)  \\
		\langle \Lambda(p',s') | \overline{s} \, i \sigma^{\mu\nu} q_\nu \gamma_5 \, b | \Lambda_b(p,s) \rangle
		=&-\overline{u}_\Lambda(p',s') \gamma_5 \Big[
		\widetilde{h}_+(q^2) \frac{q^2}{s_-}
		\left( p^\mu + p^{\prime \mu} - (m_{\Lambda_b}^2 - m_\Lambda^2)\frac{q^\mu}{q^2} \right) \notag \\
		&+ \widetilde{h}_\perp(q^2)(m_{\Lambda_b} - m_\Lambda)
		\left( \gamma^\mu + \frac{2m_\Lambda}{s_-} p^\mu - \frac{2 m_{\Lambda_b}}{s_-} p^{\prime \mu} \right)
		\Big] u_{\Lambda_b}(p,s)
	\end{align}
	with $p$ being the momentum of the $\Lambda_b$ baryon and $p^\prime$ the momentum of the $\Lambda$ baryon. $q=p-p^{\prime}$ denotes the momentum transferred. $\sigma^{\mu\nu} \equiv\frac{i}{2}(\gamma^\mu\gamma^\nu-\gamma^\nu\gamma^\mu)$ and $s_\pm \equiv(m_{\Lambda_b} \pm m_\Lambda)^2-q^2$. The above matrix elements are also usually decomposed into the first and second class form factors according to the Weinberg classification~\cite{El-Khadra:2001wco,Gutsche:2013pp},
	\begin{align}
		\langle \Lambda(p^\prime,s^\prime) | \overline{s} \,\gamma^\mu\, b | \Lambda_b(p,s) \rangle =& \overline{u}_\Lambda(p^\prime,s^\prime) \left[ f_1^V(q^2)\: \gamma^\mu - \frac{f_2^V(q^2)}{m_{\Lambda_b}} i\sigma^{\mu\nu}q_\nu + \frac{f_3^V(q^2)}{m_{\Lambda_b}} q^\mu \right] u_{\Lambda_b}(p,s),  \label{eq:WeinbergFF1} \\
		\langle \Lambda(p^\prime,s^\prime) | \overline{s} \,\gamma^\mu\gamma_5\, b | \Lambda_b(p,s) \rangle =& \overline{u}_\Lambda(p^\prime,s^\prime) \left[ f_1^A(q^2)\: \gamma^\mu - \frac{f_2^A(q^2)}{m_{\Lambda_b}} i\sigma^{\mu\nu}q_\nu + \frac{f_3^A(q^2)}{m_{\Lambda_b}} q^\mu \right]\gamma_5 u_{\Lambda_b}(p,s),  \label{eq:WeinbergFF2} \\
		\langle \Lambda(p^\prime,s^\prime) | \overline{s} \,i\sigma^{\mu\nu}q_\nu\, b | \Lambda_b(p,s) \rangle =& \overline{u}_\Lambda(p^\prime,s^\prime) \left[  \frac{f_1^{TV}(q^2)}{m_{\Lambda_b}} \left(\gamma^\mu q^2 - q^\mu \slashed{q} \right) - f_2^{TV}(q^2) i\sigma^{\mu\nu}q_\nu  \right] u_{\Lambda_b}(p,s), \label{eq:WeinbergFF3} \\
		\langle \Lambda(p^\prime,s^\prime) | \overline{s} \,i\sigma^{\mu\nu}q_\nu\,\gamma_5\, b | \Lambda_b(p,s) \rangle =& \overline{u}_\Lambda(p^\prime,s^\prime) \left[  \frac{f_1^{TA}(q^2)}{m_{\Lambda_b}} \left(\gamma^\mu q^2 - q^\mu \slashed{q} \right) - f_2^{TA}(q^2) i\sigma^{\mu\nu}q_\nu  \right]\gamma_5 u_{\Lambda_b}(p,s). \label{eq:WeinbergFF4}
	\end{align}
	The relations between these two definitions are derived as~\cite{Detmold:2016pkz},
	\begin{eqnarray}
		f_+(q^2)     &=& f_1^V(q^2) + \frac{q^2}{m_{\Lambda_b}(m_{\Lambda_b}+m_\Lambda)} f_2^V(q^2), \label{eq:FFR1} \\
		f_\perp(q^2) &=& f_1^V(q^2) + \frac{m_{\Lambda_b}+m_\Lambda}{m_{\Lambda_b}} f_2^V(q^2),  \\
		f_0(q^2)     &=& f_1^V(q^2) + \frac{q^2}{m_{\Lambda_b}(m_{\Lambda_b}-m_\Lambda)} f_3^V(q^2), \\
		g_+(q^2)     &=& f_1^A(q^2) - \frac{q^2}{m_{\Lambda_b}(m_{\Lambda_b}-m_\Lambda)} f_2^A(q^2), \\
		g_\perp(q^2) &=& f_1^A(q^2) - \frac{m_{\Lambda_b}-m_\Lambda}{m_{\Lambda_b}} f_2^A(q^2), \\
		g_0(q^2)     &=& f_1^A(q^2) - \frac{q^2}{m_{\Lambda_b}(m_{\Lambda_b}+m_\Lambda)} f_3^A(q^2), \\
		h_+(q^2)     &=& -f_2^{TV}(q^2) - \frac{m_{\Lambda_b}+m_\Lambda}{m_{\Lambda_b}} f_1^{TV}(q^2),  \\
		h_\perp(q^2) &=& -f_2^{TV}(q^2) - \frac{q^2}{m_{\Lambda_b}(m_{\Lambda_b}+m_\Lambda)} f_1^{TV}(q^2),  \\
		\widetilde{h}_+(q^2)     &=& -f_2^{TA}(q^2) + \frac{m_{\Lambda_b}-m_\Lambda}{m_{\Lambda_b}} f_1^{TA}(q^2),  \\
		\widetilde{h}_\perp(q^2) &=& -f_2^{TA}(q^2) + \frac{q^2}{m_{\Lambda_b}(m_{\Lambda_b}-m_\Lambda)} f_1^{TA}(q^2). \label{eq:FFR10}
	\end{eqnarray}

	In the framework of PQCD approach, the $\Lambda_b \to \Lambda$ transition matrix elements can be expressed as convolution of the hard kernel with the baryon LCDAs,
	\begin{eqnarray}
		\mathcal{A}\sim \Psi_{\Lambda_{b}}\left(x_{i},b_{i},\mu \right)\otimes\mathcal{H}\left(x_{i},b_{i},x^{\prime}_{i},b^{\prime}_{i},t,\mu \right)\otimes\Psi_{\Lambda}\left(x^{\prime}_{i},b^{\prime}_{i},\mu \right),
	\end{eqnarray}
	where $\Psi_{\Lambda_{b}}$ and $\Psi_{\Lambda}$ denote the LCDAs of the $\Lambda_b$ and $\Lambda$ baryons, respectively. These wave functions are nonperturbative and constitute the dominant source of theoretical uncertainty in our calculation. The hard kernel $\mathcal{H}$ is computed perturbatively from the relevant Feynman diagrams. $x_{i}$ and $x^{'}_{i}$ represent the longitudinal momentum fractions carried by quarks inside the $\Lambda_{b}$ and $\Lambda$, while $b_{i}\left(b^{'}_{i}\right)$ are the conjugate variables to the quarks' transverse momentum $k_{iT}\left(k^{'}_{iT}\right)$, respectively. The hard scale $t$ corresponds to the maximum virtuality of internal propagators in $\mathcal{H}$. Further details on the PQCD formalism and its application to baryon decays can be found in~\cite{Li:2001ay,Lu:2000hj,Han:2022srw}.
	
\begin{figure}[!htbp]
		\centering
		\includegraphics[width=1.0\textwidth]{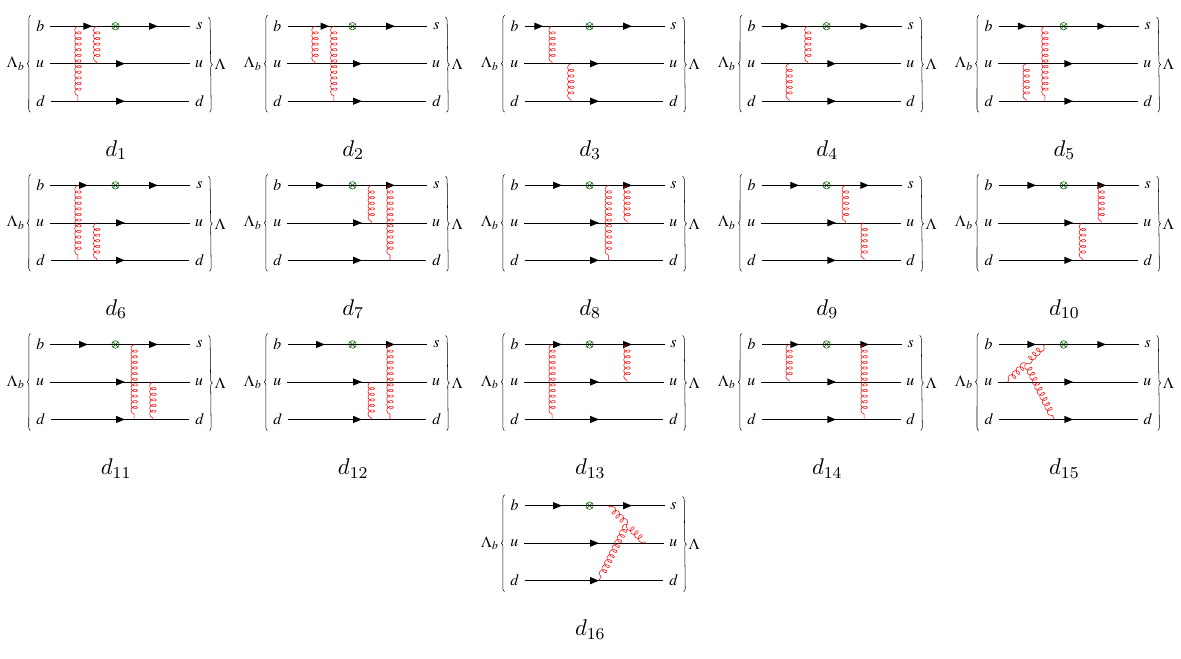}
		\captionsetup{justification=raggedright,singlelinecheck=false}
		\caption{The leading order Feynman diagrams for the $\Lambda_b\to\Lambda$ transition form factors in the PQCD approach. The green $\otimes$ denote the weak interaction currents.}
		\label{fig_all}
	\end{figure}

	In the rest frame of the $\Lambda_{b}$ baryon, the momentum of the $\Lambda_{b}$ baryon and $\Lambda$ baryon are parameterized in the light-cone coordinate as,
	
	\begin{eqnarray}
		p=\frac{m_{\Lambda_b}}{\sqrt{2}}(1,1,\vec{\textbf{0}}),\quad p^\prime=\frac{m_{\Lambda_b}}{\sqrt{2}}(\eta_1,\eta_2,\vec{\textbf{0}}),
	\end{eqnarray}
	with
	\begin{eqnarray}
		\eta_1=\frac{M^{2}_{\Lambda_{b}}+m^{2}_{\Lambda}-q^{2}\pm\sqrt{\left(M^{2}_{\Lambda_{b}}+m^{2}_{\Lambda}-q^{2}\right)^{2}-4M^{2}_{\Lambda_{b}}m^{2}_{\Lambda}}}{2M^{2}_{\Lambda_{b}}},\nonumber\\
		\eta_2=\frac{M^{2}_{\Lambda_{b}}+m^{2}_{\Lambda}-q^{2}\mp\sqrt{\left(M^{2}_{\Lambda_{b}}+m^{2}_{\Lambda}-q^{2}\right)^{2}-4M^{2}_{\Lambda_{b}}m^{2}_{\Lambda}}}{2M^{2}_{\Lambda_{b}}},
	\end{eqnarray}
where $M_{\Lambda_{b}}$ and $m_{\Lambda}$ denote the masses of the $\Lambda_b$ and $\Lambda$, respectively. The momentum for quarks with the transverse momentum retained are parameterized as,
	\begin{align}
		k_1 &= \left(\frac{m_{\Lambda_b}}{\sqrt{2}}, \frac{x_1 m_{\Lambda_b}}{\sqrt{2}}, \vec{\textbf{k}}_{1T}\right), & k_1^\prime &= \left(x_1^\prime \frac{\eta_1 m_{\Lambda_b}}{\sqrt{2}}, 0, \vec{\textbf{k}}_{1T}^\prime\right), \nonumber \\
		k_2 &= \left(0, \frac{x_2 m_{\Lambda_b}}{\sqrt{2}}, \vec{\textbf{k}}_{2T}\right), & k_2^\prime &= \left(x_2^\prime \frac{\eta_1 m_{\Lambda_b}}{\sqrt{2}}, 0, \vec{\textbf{k}}_{2T}^\prime\right), \nonumber \\
		k_3 &= \left(0, \frac{x_3 m_{\Lambda_b}}{\sqrt{2}}, \vec{\textbf{k}}_{3T}\right), & k_3^\prime &= \left(x_3^\prime \frac{\eta_1 m_{\Lambda_b}}{\sqrt{2}}, 0, \vec{\textbf{k}}_{3T}^\prime\right).
	\end{align}
	$k_1$ and $k_1^\prime$ represent the momenta of the $b$ and $s$ quarks, respectively, while $k_2$ ($k_3$) and $k_2^\prime$ ($k_3^\prime$) denote the momenta of the spectator $u$ ($d$) quarks, respectively. In the $\Lambda_{b}$ baryon, $b$ quark carries the majority of the longitudinal momentum, $x_{1}\sim {\cal O}\left(m_{b}^{2}/M^{2}_{\Lambda_{b}}\right)$, where $m_b$ is the mass of b quark. $u, d$ quarks carry small momentum and are treated as soft degrees of freedom. As illustrated in Fig.~\ref{fig_all}, at least two hard gluon exchanges are required to transfer momentum to these soft quarks to form the energetic $\Lambda$ baryon in the final state. This implies that the $\Lambda_b \to \Lambda$ transition amplitude in the PQCD approach starts at $\mathcal{O}(\alpha_s^2)$, which constitutes the leading-order contribution.

	\section{Light-cone distribution amplitudes}\label{sec:LCDAs}
	As stated earlier, the essential inputs for calculating the $\Lambda_{b}\to\Lambda$ transition form factors in the PQCD approach are the LCDAs of the baryons.
	The LCDAs of the $\Lambda_b$ baryon are defined by the matrix elements of non-local operators between the vacuum and the $\Lambda_b$ baryon state. The corresponding Lorentz structures in momentum space take the form~\cite{Wang:2015ndk,Bell:2013tfa,Ball:2008fw}
	{\small
		\begin{equation}
			\begin{split}
				(Y_{\Lambda_b})_{\alpha\beta\gamma}(x_i,\mu)&\equiv \frac{1}{2N_c}\int \prod_{l=2}^{3} \frac{dz_l^-d\bm{z}_l}{(2\pi)^3}e^{ik_l\cdot z_l} \epsilon^{ijk}\langle 0|T[b_\alpha^i(0) u_\beta^j(z_2)d_\gamma^k(z_3)]|\Lambda_b\rangle\\
				&=\frac{1}{8N_c}\Big\{f_{\Lambda_b}^{(1)}(\mu)[M_1(x_2,x_3)\gamma_5C^T]_{\gamma\beta}+f_{\Lambda_b}^{(2)}(\mu)[M_2(x_2,x_3)\gamma_5C^T]_{\gamma\beta}\Big\}[\Lambda_b(p)]_\alpha,\label{pro}
			\end{split}
	\end{equation}}
	where $\alpha,\beta,\gamma$ are the spinor indices, $i,j,k$ are the color indices and $N_{c}$ is the number of colors. $C^{T}$ represents the transpose of the charge conjugate matrix, and $\Lambda_{b}\left(p\right)$ represrnts the spinor of the $\Lambda_{b}$ baryon. $f_{\Lambda_{b}}^{\left(1\right)}$ and $f_{\Lambda_{b}}^{\left(2\right)}$ are normalized constants, here we choose $f_{\Lambda_{b}}^{\left(1\right)}\approx f_{\Lambda_{b}}^{\left(2\right)}\equiv f_{\Lambda_{b}}=0.031\pm0.005GeV^{3}$~\cite{Han:2022srw,Rui:2022sdc}. The remaining parts of the chiral-even projector $M_{1}$ and the chiral-odd projector $M_{2}$ are expressed as,
	\begin{eqnarray}
		M_1(x_2,x_3)&=&\frac{\slashed{\bar{n}}\slashed{n}}{4}\psi_3^{+-}(x_2,x_3)+\frac{\slashed{n}\slashed{\bar{n}}}{4}\psi_3^{-+}(x_2,x_3),\nonumber\\
		M_2(x_2,x_3)&=&\frac{\slashed{n}}{\sqrt{2}}\psi_2(x_2,x_3)+\frac{\slashed{\bar{n}}}{\sqrt{2}}\psi_4(x_2,x_3),
	\end{eqnarray}
	where $n$ and $\bar{n}$ are two light-cone vectors, $n=\left(1,0,\vec{\textbf{0}}_T\right)$, $\bar{n}=\left(0,1,\vec{\textbf{0}}_T\right)$, satisfying $n\cdot\bar{n}=1$. Several theoretical models for the $\Lambda_{b}$ baryon LCDA functions $\Psi_{2},\Psi_{3}^{+-},\Psi_{3}^{-+},\Psi_{4}$ have been proposed in~\cite{Ball:2008fw,Bell:2013tfa,Ali:2012zza}. In this work we adopt the exponential mode from~\cite{Bell:2013tfa} which is in a simple form and considered to be good enough in current stage~\cite{Han:2022srw,Rui:2022sdc}. The exponential mode is given by~\cite{Bell:2013tfa}
	\begin{eqnarray}\label{lb_function}
		\Psi_2(x_2,x_3)&=&     x_2x_3\frac{M_{\Lambda_{b}}^4}{\omega_0^4}e^{-\frac{(x_2+x_3)M_{\Lambda_{b}}}{\omega_0}},\nonumber\\
		\Psi_3^{+-}(x_2,x_3)&=&2x_2\frac{M_{\Lambda_{b}}^3}  {\omega_0^3}e^{-\frac{(x_2+x_3)M_{\Lambda_{b}}}{\omega_0}},\nonumber\\
		\Psi_3^{-+}(x_2,x_3)&=&2x_3\frac{M_{\Lambda_{b}}^3}  {\omega_0^3}e^{-\frac{(x_2+x_3)M_{\Lambda_{b}}}{\omega_0}},\nonumber\\
		\Psi_4(x_2,x_3)&=&     \frac{ M_{\Lambda_{b}}^2}     {\omega_0^2}e^{-\frac{(x_2+x_3)M_{\Lambda_{b}}}{\omega_0}},
	\end{eqnarray}
	where the parameter $\omega_0$ denotes the average energy of the two light quarks and is chosen to be $0.7\pm0.1$GeV in the PQCD approach as suggested by the analyses of the $\Lambda_b\to p\pi^-,pK^-$ decays~\cite{Han:2024kgz,Han:2025tvc}.
	
	
	In the SU(3) flavor symmetry limit, the LCDAs for an outgoing final-state $\Lambda$ baryon up to twist-6 have been defined through the corresponding momentum-space projector in~\cite{Braun:2000kw,Liu:2014uha,Wein:2015oqa},	
	{\small
	\begin{align}
				(\overline{Y}_{\Lambda})_{\alpha\beta\gamma}(x_i^\prime,\mu)=&\frac{1}{2\sqrt{2}N_c}\int \prod_{l=2}^{3} \frac{dz_l^-d\bm{z}_l}{(2\pi)^3}e^{ik_l\cdot z_l} \epsilon^{ijk}\langle \Lambda(p^\prime) |\epsilon^{ijk} {\bar{u}}_\alpha^i(0) {\bar{d}}_\beta^j(z_2) {\bar{s}}_\gamma^k(z_3) |0\rangle\ \nonumber\\
				=&\frac{-1}{8\sqrt{2}N_c}\Big\{
				S_1 m_{\Lambda} C_{\beta\alpha} (\bar{\Lambda}^+ \gamma_5)_\gamma + S_2 m_{\Lambda} C_{\beta\alpha} (\bar{\Lambda}^- \gamma_5)_\gamma + P_1 m_{\Lambda} (C\gamma_5)_{\beta\alpha} \bar{\Lambda}^+_\gamma\nonumber\\
				&+ P_2 m_{\Lambda} (C\gamma_5)_{\beta\alpha} \bar{\Lambda}^-_\gamma+ V_1 (C\slashed{\mathcal{P}}^{\prime})_{\beta\alpha} (\bar{\Lambda}^+\gamma_5)_\gamma + V_2 (C\slashed{\mathcal{P}}^{\prime})_{\beta\alpha} (\bar{\Lambda}^-\gamma_5)_\gamma\nonumber\\
				&+ V_3 \frac{m_{\Lambda}}{2} (C\gamma_\perp)_{\beta\alpha}(\bar{\Lambda}^+\gamma_5\gamma^\perp)_\gamma+ V_4 \frac{m_{\Lambda}}{2} (C\gamma_\perp)_{\beta\alpha}(\bar{\Lambda}^-\gamma_5\gamma^\perp)_\gamma \nonumber\\
				&+ V_5\frac{m_{\Lambda}^2}{2\mathcal{P}^{\prime}z} (C\slashed{z})_{\beta\alpha}(\bar{\Lambda}^+\gamma_5)_\gamma+ V_6\frac{m_{\Lambda}^2}{2\mathcal{P}^{\prime}z} (C\slashed{z})_{\beta\alpha}(\bar{\Lambda}^-\gamma_5)_\gamma\nonumber\\
				&+ A_1 (C\gamma_5\slashed{\mathcal{P}}^{\prime})_{\beta\alpha} (\bar{\Lambda}^+)_\gamma+ A_2 (C\gamma_5\slashed{\mathcal{P}}^{\prime})_{\beta\alpha} (\bar{\Lambda}^-)_\gamma+ A_3 \frac{m_{\Lambda}}{2} (C\gamma_5\gamma_\perp)_{\beta\alpha}(\bar{\Lambda}^+\gamma^\perp)_\gamma\nonumber\\
				&+ A_4 \frac{m_{\Lambda}}{2} (C\gamma_5\gamma_\perp)_{\beta\alpha}(\bar{\Lambda}^-\gamma^\perp)_\gamma+ A_5\frac{m_{\Lambda}^2}{2\mathcal{P}^{\prime}z} (C\gamma_5\slashed{z})_{\beta\alpha}(\bar{\Lambda}^+)_\gamma \nonumber\\
				&+ A_6\frac{m_{\Lambda}^2}{2\mathcal{P}^{\prime}z} (C\gamma_5\slashed{z})_{\beta\alpha}(\bar{\Lambda}^-)_\gamma- T_1 (iC\sigma_{\perp \mathcal{P}^{\prime}})_{\beta\alpha}(\bar{\Lambda}^+\gamma_5\gamma^\perp)_\gamma\nonumber\\
				&- T_2 (iC\sigma_{\perp \mathcal{P}^{\prime}})_{\beta\alpha}(\bar{\Lambda}^-\gamma_5\gamma^\perp)_\gamma - T_3 \frac{m_{\Lambda}}{\mathcal{P}^{\prime}z}(iC\sigma_{\mathcal{P}^{\prime}z})_{\beta\alpha}(\bar{\Lambda}^+\gamma_5)_\gamma\nonumber\\
				&- T_4 \frac{m_{\Lambda}}{\mathcal{P}^{\prime}z}(iC\sigma_{z\mathcal{P}^{\prime}})_{\beta\alpha}(\bar{\Lambda}^-\gamma_5)_\gamma - T_5\frac{m_{\Lambda}^2}{2\mathcal{P}^{\prime}z}(iC\sigma_{\perp z})_{\beta\alpha}(\bar{\Lambda}^+\gamma_5\gamma^\perp)_\gamma \nonumber\\
				&- T_6\frac{m_{\Lambda}^2}{2\mathcal{P}^{\prime}z}(iC\sigma_{\perp z})_{\beta\alpha}(\bar{\Lambda}^-\gamma_5\gamma^\perp)_\gamma+ T_7\frac{m_{\Lambda}}{2}(C\sigma_{\perp\perp^\prime})_{\beta\alpha}(\bar{\Lambda}^+\gamma_5\sigma^{\perp\perp^\prime})_\gamma \nonumber\\
				&+ T_8\frac{m_{\Lambda}}{2}(C\sigma_{\perp\perp^\prime})_{\beta\alpha}(\bar{\Lambda}^-\gamma_5\sigma^{\perp\perp^\prime})_\gamma
				\Big\},\label{matrix}
\end{align}
	The light-like vector $\mathcal{P^{\prime}}$ is defined, by the $\Lambda$ baryon momentum $p^\prime$ and the light-like vector $z$, as
	\begin{equation}
		\mathcal{P^{\prime}}_\mu=p^{\prime}_\mu-\frac{1}{2}z_\mu\frac{m_{\Lambda}^2}{\mathcal{P^{\prime}}\cdot z},
	\end{equation}
	satisfying $\mathcal{P^{\prime}}\cdot z\sim 1$, and $\mathcal{P^{\prime}}\to p^{\prime}$ in the limit of $m_{\Lambda}\to 0$.
	In the definition of Eq.~(\ref{matrix}), the spinor of the $\Lambda$ baryon has been decomposed into the ``large'' component, $\bar{\Lambda}^+=\bar{\Lambda}\slashed{z}\slashed{\mathcal{P}}^\prime/2$, and ``small'' component, $\bar{\Lambda}^-=\bar{\Lambda}\slashed{\mathcal{P}}^\prime\slashed{z}/2$.
	The shorthand notations have been introduced as $\sigma_{\mathcal{P}^{\prime}z}=\sigma^{\mu\nu}\mathcal{P}^{\prime}_\mu z_\nu$, $\gamma_\perp\gamma^\perp=\gamma^\mu g_{\mu\nu}^\perp\gamma^\nu$ with $g_{\mu\nu}^\perp=g_{\mu\nu}-(\mathcal{P}^{\prime}_\mu z_\nu+z_\mu \mathcal{P}^{\prime}_\nu)/\mathcal{P}^{\prime}z$. The terms $V_1$, $A_1$ and $T_1$, together with the Lorentz structures are classified as leading twist DAs. The other terms in Eq.~(\ref{matrix}) are higher twist DAs with definite twist classification.

    Refs.~\cite{Bali:2015ykx,RQCD:2019hps} systematically research the normalization constants and first moments of the leading twist distribution amplitudes of the baryon octet in the lattice QCD. Leading twist DAs can be expanded in a set of orthogonal polynomials under conformal partial wave expansion as,
    \begin{equation}
    \begin{split}
        V_1(x_1,x_2,x_3)&=120x_1x_2x_3(x_1-x_2)\left(\frac{21\sqrt{6}}{4}\phi_{11}-\frac{7\sqrt{6}}{4}\phi_{10}\right)\\
        A_1(x_1,x_2,x_3)&=120x_1x_2x_3\left(-f_{\Lambda}+\frac{7\sqrt{6}}{4}(\phi_{11}+\phi_{10})(x_1+x_2)-\frac{14\sqrt{6}}{4}x_3(\phi_{11}+\phi_{10})\right)\\
        T_1(x_1,x_2,x_3)&=120x_1x_2x_3\pi_{10}\frac{7\sqrt{6}}{2}(x_1-x_2).
    \end{split}
    \end{equation}
    The leading contributions $120x_1x_2x_3f_{\Lambda}$ is usually referred to  as the asymptotic DAs with the normalization coefficient $f_{\Lambda}$. Higher-order coefficients $\phi_{10}$, $\phi_{11}$ and $\pi_{10}$ are the first moments and usually referred to as shape parameters. Those parameters, rescaled to $1GeV$, are obtained in the the continuum limit, $f_{\Lambda}=6.20^{+0.11}_{-0.09}\times 10^{-3}GeV^2$, $\phi_{11}=0.249^{+0.058}_{-0.058}\times 10^{-3}GeV^2$, $\phi_{10}=0.719^{+0.040}_{-0.044}\times 10^{-3}GeV^2$ and $\pi_{10}=0.252^{+0.040}_{-0.033}\times 10^{-3}GeV^2$ \cite{RQCD:2019hps}.
    
   For the high-twist DAs $V_i,A_i,T_i,S_i$ and $P_i$ of the $\Lambda$ baryon, we employ the parameterization given in Ref.~\cite{Liu:2009mb}, which is directly extended from the nucleon LCDAs in Ref.~\cite{Braun:2000kw} under the constraints of the equations of motion and general symmetry,
    \begin{eqnarray}
    S_1(x_1,x_2,x_3)&=&6x_3(1-x_3)(\xi_4^0+\xi_4^{'0})\,,\hspace{1.5cm}P_1(x_1,x_2,x_3)=6x_3(1-x_3)(\xi_4^0-\xi_4^{'0})\,,\nonumber\\
    V_2(x_1,x_2,x_3)&=&0\,,\hspace{5.0cm}A_2(x_1,x_2,x_3)=-24x_1x_2\phi_4^0\,,\nonumber\\
    V_3(x_1,x_2,x_3)&=&12(x_1-x_2)x_3\psi_4^0\,,\hspace{2.3cm}A_3(x_1,x_2,x_3)=-12x_3(1-x_3)\psi_4^0\,,\nonumber\\
    T_2(x_1,x_2,x_3)&=&0\,,\hspace{5.0cm}T_3(x_1,x_2,x_3)=6(x_2-x_1)x_3(-\xi_4^0+\xi_4^{'0})\,,\nonumber\\
    T_7(x_1,x_2,x_3)&=&-6(x_1-x_2)x_3(\xi_4^0+\xi_4^{'0})\,\nonumber\\
    S_2(x_1,x_2,x_3)&=&\frac32(x_1+x_2)(\xi_5^0+\xi_5^{'0})\,,\hspace{1.5cm}P_2(x_1,x_2,x_3)=\frac32(x_1+x_2)(\xi_5^0-\xi_5^{'0})\,,\nonumber\\
    V_4(x_1,x_2,x_3)&=&3(x_2-x_1)\psi_5^0\,,\hspace{2.8cm}A_4(x_1,x_2,x_3)=-3(1-x_3)\psi_5^0\,,\nonumber\\
    V_5(x_1,x_2,x_3)&=&0\,,\hspace{4.9cm}A_5(x_1,x_2,x_3)=-6x_3\phi_5^0\,,\nonumber\\
    T_4(x_1,x_2,x_3)&=&-\frac32(x_1-x_2)(\xi_5^0+\xi_5^{'0})\,,\hspace{1.3cm}T_5(x_1,x_2,x_3)=0\,,\nonumber\\
    T_8(x_1,x_2,x_3)&=&-\frac32(x_1-x_2)(\xi_5^0-\xi_5^{'0})\,\nonumber\\
    V_6(x_1,x_2,x_3)&=&0\,,\hspace{5cm}A_6(x_1,x_2,x_3)=-2\phi_6^0\,,\nonumber\\
    T_6(x_1,x_2,x_3)&=&0\,.
    \end{eqnarray}
    The recently developed large-momentum effective theory (LaMET) enables extraction of fully $x$-dependent distribution amplitudes from first principles on the lattice \cite{LatticeParton:2024vck,LPC:2025jvd}, including the contribution of higher moments, which could provide more reliable inputs for future precision calculations of baryons.
     
    The parameters $\phi_i^0$, $\psi_i^0$, $\xi_i^0$ and $\xi_i^{\prime 0}$ in the above DAs connect to the $f_{\Lambda}$ and three new P-wave (high twist) normalization constants $\lambda_{1,2,3}$ as~\cite{Liu:2009mb,Bali:2015ykx},
	\begin{eqnarray}\label{lb_parameter}
    \phi_6^0&=&f_\Lambda,\hspace{4.7cm}\phi_4^0=\phi_5^0=-\frac{1}{2}(f_\Lambda-\frac{1}{\sqrt{6}}\lambda_1),\nonumber\\
		\psi_4^0&=&\psi_5^0=\frac12(f_\Lambda+\frac{1}{\sqrt{6}}\lambda_1),\hspace{1.6cm}\xi_4^0=\xi_5^0=-\frac{\lambda_2}{2\sqrt{6}},\nonumber\\
		\xi_4^{'0}&=&\xi_5^{'0}=-\frac{\lambda_3}{\sqrt{6}}.
	\end{eqnarray}
    $\lambda_{1,2,3}$ can be defined as matrix elements of local three-quark twist-four operators without derivatives~\cite{Braun:2000kw,Wein:2015oqa},
    \begin{equation}
    \begin{split}
        \langle 0|\left( u^T(0)C\gamma^\mu\gamma_5d(0)\right)\gamma_\mu s(0)|\Lambda(p,\lambda)\rangle=&\frac{-1}{\sqrt{6}}\lambda_1m_{\Lambda}u_{\Lambda}(p,\lambda),\\
        \langle 0|\left( u^T(0)Cd(0)\right)\gamma_5 s(0)|\Lambda(p,\lambda)\rangle=&\frac{-1}{4\sqrt{6}}(\lambda_2+2\lambda_3)m_{\Lambda}u_{\Lambda}(p,\lambda),\\
        \langle 0|\left( u^T(0)C\gamma_5d(0)\right) s(0)|\Lambda(p,\lambda)\rangle=&\frac{-1}{4\sqrt{6}}(\lambda_2-2\lambda_3)m_{\Lambda}u_{\Lambda}(p,\lambda).
    \end{split}
    \end{equation}
    The explicit values from lattice QCD calculation for the parameters $\lambda_{1,2,3}$ are $\lambda_1=(-37.6\pm 3.5)\times 10^{-3}GeV^2$, $\lambda_2=(88.3\pm 4.6)\times 10^{-3}GeV^2$ and $\lambda_3=(-46.7\pm 4.5)\times 10^{-3}GeV^2$~\cite{RQCD:2019hps}.

	\section {numerical results}\label{sec:FFs}
	The PQCD factorization formulas for all relevant Feynman diagrams are provided in Appendix~\ref{appendix}. We present our numerical results for the $\Lambda_b \to \Lambda$ form factors at $q^2=0\:\text{GeV}^2$ in Table~\ref{table_1}, and compare with those from the lattice QCD~\cite{Detmold:2016pkz}. The uncertainties in this work arise from the parameters describing the $\Lambda_b$ and $\Lambda$ baryon LCDAs. Further improvements in the determination of baryon LCDAs will help reduce theoretical uncertainties in future predictions. 
    
    As shown in Table~\ref{table_1}, our results are consistent with those from the lattice QCD within uncertainties and satisfy the expected endpoint relations $f_0(0)=f_+(0)$ and $g_0(0)=g_+(0)$.
    For further comparison with the form factors from other approaches, we convert the obtained helicity form factors in Table~\ref{table_1} to Weinberg form factors in Table~\ref{table_compare}. Our predictions are consistent with expectations from heavy-quark effective theory (HQET), which predicts for heavy-to-light baryon transitions form factors at $q^2=0$, $f_1^V(0) \approx f_1^A(0) \gg f_{2,3}^{V,A}(0)$, $f_2^{TV}(0) \approx f_2^{TA}(0) \gg f_{1}^{TV,TA}(0)$.
	\begin{table}[htbp]
		\centering
		\renewcommand{\arraystretch}{1.1}
        \captionsetup{justification=raggedright,singlelinecheck=false}
		\caption{Form factors of the $\Lambda_b \to \Lambda$ transition obtained in the PQCD and lattice QCD in helicity basis at $q^2=0GeV^2$.}
		\label{table_1}
		\begin{tabular*}{120mm}{c@{\extracolsep{\fill}}cc}
			\hline\hline
			& this work & lattice QCD~\cite{Detmold:2016pkz}\\
            \hline
			$f_{+}(0)$ & $0.149\pm0.035$ & $0.160\pm0.024$\\
			$f_{\perp}(0)$   & $0.185\pm0.043$ & $0.207\pm0.037$\\
			$f_0(0)$ & $0.149\pm0.036$   & $0.156\pm0.035$\\
			$g_{+}(0)$   & $0.159\pm0.038$  & $0.112\pm0.028$\\
			$g_{\perp}(0)$   & $0.139\pm0.034$  & $0.095\pm0.035$\\
			$g_0(0)$ & $0.159\pm0.038$  & $0.166\pm0.023$\\
			$h_{+}(0)$   & $0.185\pm0.043$ & $0.236\pm0.041$\\
			$h_{\perp}(0)$ & $0.154\pm0.037$ & $0.166\pm0.022$\\
			$\widetilde{h}_{+}(0)$ & $0.138\pm0.034$  & $0.163\pm0.029$\\
			$\widetilde{h}_{\perp}(0)$ & $0.154\pm0.036$   & $0.156\pm0.027$\\
			\hline\hline
		\end{tabular*}
	\end{table}
	

	
    \begin{table}[htbp]
    \scriptsize
    \renewcommand{\arraystretch}{1.1}
    \captionsetup{justification=raggedright,singlelinecheck=false}
    \caption{Theoretical predictions for the $\Lambda_b\rightarrow \Lambda$ form factors in Weinberg basis at $q^2=0$. The symbol $\cdots$ denotes that there is no relevant result in literature.}
    \label{table_compare}
    \begin{tabular*}{165mm}{c@{\extracolsep{\fill}}cccccc}
        \hline\hline
        & $f_1^V(0)$ & $f_2^V(0)$ & $f_3^V(0)$ & $f_1^A(0)$ & $f_2^A(0)$ & $f_3^A(0)$\\
        \hline
        this work & $0.149\pm0.036$ & $0.030\pm0.006$ & $-0.015\pm0.003$ & $0.159\pm0.039$ & $0.025\pm0.006$ & $-0.020\pm0.004$ \\			
        Lattice QCD\cite{Detmold:2016pkz} & $0.158\pm0.004$ & $0.041\pm0.005$ & $-0.017\pm0.009$ & $0.137\pm0.004$ & $0.055\pm0.008$ & $-0.117\pm0.009$ \\
        LCSR\cite{Wang:2015ndk} & $0.18$ & $0.017$ & $\cdots$ & $0.18$ & $-0.025$ & $\cdots$ \\
        LCSR\cite{Aliev:2010uy} & $0.322\pm0.112$ & $0.062\pm0.022$ & $-0.084\pm0.028$ & $0.318\pm0.110$ & $0.073\pm0.022$ & $-0.079\pm0.028$ \\
        relativistic QM\cite{Faustov:2017wbh} & $0.208$ & $0.302$ & $0.026$ & $0.125$ & $-0.003$ & $-0.083$ \\
        covariant QM\cite{Gutsche:2013oea} & $0.107$ & $0.043$ & $0.003$ & $0.104$ & $0.003$ & $-0.052$ \\
        QCDSR\cite{Huang:1998ek} & $0.446$ & $0.073$ & $-0.073$ & $0.446$ & $0.073$ & $-0.073$ \\
        PQCD\cite{Rui:2022sdc} & $0.095^{+0.057+0.018}_{-0.029-0.021}$ & $\cdots$ & $\cdots$ & $0.104^{+0.060+0.016}_{-0.033-0.020}$ & $\cdots$ & $\cdots$ \\
        CLFQM\cite{Zhu:2018jet} & $0.131_{-0.018-0.008}^{+0.016+0.008}$ & $0.048_{-0.008-0.002}^{+0.008+0.002}$ & $-0.027_{-0.006-0.000}^{+0.006+0.000}$ & $0.132_{-0.017-0.009}^{+0.016+0.008}$ & $0.024_{-0.005-0.001}^{+0.005+0.001}$ & $-0.052_{-0.008-0.001}^{+0.008+0.001}$  \\
        NRQM\cite{Mott:2011cx} & $0.025$ & $\cdots$ & $\cdots$ & $0.028$ & $\cdots$ & $\cdots$ \\
        LFQM\cite{Wei:2009np} & $0.108$ & $0.031$ & $-0.031$ & $0.106$ & $0.006$ & $-0.006$ \\
        GFA\cite{Mohanta:2000nk} & $0.061$ & $0.025$ & $-0.008$ & $0.107$ & $0.014$ & $-0.043$ \\
        NRQM\cite{Cheng:1996cs} & $0.062$ & $0.025$ & $-0.008$ & $0.108$ & $0.014$ & $-0.043$ \\
        \hline
        & $f^{TV}_1(0)$ & $f^{TV}_2(0)$ & $f^{TA}_1(0)$ & $f^{TA}_2(0)$ &\\
        \hline
        this work& $-0.015\pm0.006$ & $-0.154\pm0.037$ & $-0.019\pm0.004$ & $-0.154\pm0.036$ \\
        Lattice QCD\cite{Detmold:2016pkz} & $-0.059\pm0.006$ & $-0.166\pm0.005$ & $0.009\pm0.008$ & $-0.156\pm0.005$ \\
        LCSR\cite{Wang:2015ndk} & $-0.025$ & $-0.18$ & $0.037$ & $-0.18$ \\
        LCSR\cite{Aliev:2010uy} & $-0.070\pm0.022$ & $-0.295\pm0.105$ & $-0.091\pm0.032$ & $-0.294\pm0.105$ \\
        relativistic QM\cite{Faustov:2017wbh} & $-0.029$ & $-0.153$ & $0.029$ & $-0.153$ \\
        covariant QM\cite{Gutsche:2013oea} & $-0.043$ & $-0.105$ & $0.003$ & $-0.105$ \\
        QCDSR\cite{Huang:1998ek} & $-0.076$ & $-0.446$ & $-0.083$ & $-0.446$ \\
        \hline
        \hline
    \end{tabular*}
    \end{table}
	

	
    Contributions to the form factors $f_1^{V,A}(0)$ from each Feynman diagram are presented in Table~\ref{tab:forme}, which reveals that the diagrams $d_7$ and $d_{8}$ dominate.
    We analyze the contributions of each twist-term combination for these two diagrams in Table~\ref{tab:twist-contribution}, which shows that the $\Lambda_b$ twist-4 and $\Lambda$ twist-5 combination accounts for the dominant numerical results in both $d_7$ and $d_8$. These results clearly demonstrate the essential role of higher-power effects in baryon decays at the $b$-quark mass scale.
	\begin{table}[htbp]
		\centering
		\small
		\renewcommand{\arraystretch}{1.1}
		\captionsetup{justification=raggedright,singlelinecheck=false}
		\caption{All diagram results for the form factors $f_1^V$ and $f_1^A$ at $q^2=0$ of the $\Lambda_b\rightarrow \Lambda$ transition.}
		\label{tab:forme}
		\begin{tabular*}{165mm}{c@{\extracolsep{\fill}}cccccccc}
			\hline\hline
			& $d_1$ & $d_{2}$ & $d_{3}$ & $d_{4}$ & $d_{5}$ & $d_{6}$ & $d_{7}$ & $d_{8}$ \\ 
			\hline
			$f_1^V(0)$ & $-4.7\times10^{-4}$ & $-4.6\times10^{-4}$ & $6.5\times10^{-3}$ & $7.1\times10^{-4}$ & $-3.5\times10^{-3}$& $-4.2\times10^{-3}$ & 0.107 & 0.093 \\
			$f_1^A(0)$ & $-3.4\times10^{-4}$ & $2.1\times10^{-5}$ & $5.9\times10^{-3}$ & $5.7\times10^{-4}$ & $3.5\times10^{-3}$& $-3.6\times10^{-3}$ & 0.106 & 0.092\\ 
			\hline
			& $d_{9}$ & $d_{10}$ & $d_{11}$ & $d_{12}$&$d_{13}$&$d_{14}$&total&   \\ 
			\hline
			$f_1^V(0)$	& $-5.8\times10^{-3}$ & $-4.6\times10^{-4}$ & 0.021 & $-1.1\times10^{-3}$ & -0.028 & -0.034 & 0.149& \\
			$f_1^A(0)$ & $-5.9\times10^{-3}$ & $-3.8\times10^{-4}$ & 0.021 & $-1.1\times10^{-3}$ & -0.024 & -0.029 & 0.159  &\\	 		
			\hline\hline
		\end{tabular*}
	\end{table}
	
	

	\begin{table}[htbp]
		\small
		\renewcommand\arraystretch{1.1}
			\captionsetup{justification=raggedright,singlelinecheck=false}
		\caption{The contribution of the diagrams $d_7,d_8,d_{13}$ and $d_{14}$ to the form factor $f_1^V(0)$.}
		\centering
		\begin{tabular*}{165mm}{c@{\extracolsep{\fill}}|cccc|cccc}
			\hline
			\hline
			& \multicolumn{4}{c|}{$d_7$} & \multicolumn{4}{c}{$d_8$} \\
			\hline
			& $\Lambda$ twist-3 & $\Lambda$ twist-4 & $\Lambda$ twist-5 & $\Lambda$ twist-6  & $\Lambda$ twist-3 & $\Lambda$ twist-4 & $\Lambda$ twist-5 & $\Lambda$ twist-6 \\
			\hline
			$\Lambda_{b}$ twist-2 & $\sim 0$ & $\sim 0$ & $\sim 0$ & $\sim 0$ & $\sim 0$ & $\sim 0$ & $\sim 0$ & $\sim 0$ \\
			\hline
			$\Lambda_{b}$ twist-3 & $\sim 0$ & $\sim 0$ & $\sim 0$ & $\sim 0$ & $\sim 0$ & $\sim 0$ & $\sim 0$ & $\sim 0$ \\
			\hline
			$\Lambda_{b}$ twist-4 & -0.017 & $\sim 0$ & 0.118 & $\sim 0$ & -0.015 & $\sim 0$ & 0.103 & $\sim 0$ \\
			\hline
			\hline
		\end{tabular*}
		\label{tab:twist-contribution}
	\end{table}

	Recalling that the form factors are expressed as the convolution of the perturbatively calculated hard kernels and the non-perturbative baryon LCDAs. We list the numerators $h_{ij}$ of the hard kernels for the diagram $d_7$ contributing to the form factor $f_1^V(0)$ in Table~\ref{Tab_d7f}, classified by different twist combination. $i=2,3,4$ stands for the $\Lambda_b$ baryon LCDA and $j=3,4,5,6$ denotes the $\Lambda$ baryon LCDA.
	The mass ratio, $r=m_{\Lambda}/M_{\Lambda_{b}}$, reflects the $1/m_b$ power suppression associated with the high-twist $\Lambda$ baryon LCDAs. 
	By comparing the $h_{43}$ and $ h_{45}$ terms in Table~\ref{Tab_d7f}, it is found that $h_{45}$ is power suppressed by factor of $r^2$ compared to $h_{43}$. However, $h_{43}$ includes an additional momentum fraction $x_{3}$.
	
	Fig.~\ref{lcda_lbe} displays the dependence of the $\Lambda_b$ baryon LCDAs on the momentum fractions $x_1$, $x_2$ and $x_3$ in the exponential model. The twist-4 $\Lambda_b$ LCDA $\Psi_4$ shows a marked peak where $x_1 \approx 1$ and $x_{2,3}$ approach zero, clearly indicating that the b-quark carries most of the baryon's momentum.
	The presence of the momentum fraction $x_3$ in $h_{43}$ closing to zero significantly suppresses the contribution from the endpoint peak region of the twist-4 $\Lambda_b$ LCDA $\Psi_4$.
	This pattern demonstrates that high-power contributions to the baryon transition form factors are significantly enhanced near the endpoint region, emerging as the dominant effects, which is in agreement with the earlier findings in Ref.~\cite{Wang:2011uv,Han:2022srw}.
	\begin{table}[tbhp]
		\centering
		\footnotesize
		\renewcommand{\arraystretch}{1.1}
		\caption{Numerators $h_{ij}$ of the hard kernels in the factorization formula for the form factor $f_1^V(0)$ with $i=2,3^{+-},3^{-+},4$ and $j=3,4,5,6$. An overall coefficient $\mathcal{C}m_{\Lambda_b}^3f_{\Lambda_b} /(128N_c^2)$ with the color factor $\mathcal{C}=8/3$ is implicit for each entry.}
		\begin{tabular*}{165mm}{l@{\extracolsep{\fill}}cccc}
			\hline
			\hline
			$h_{ij}$ & $\Lambda$ twist-3 & $\Lambda$ twist-4 & $\Lambda$ twist-5 & $\Lambda$ twist-6 \\
			\hline
			$\Lambda_{b}$ twist-2 & $\sim 0$ & $\sim 0$ & $-8r^{2}x_{3}\Psi_{2}\left(A_{4}+A_{5}+V_{5}-V_{4}\right)$ & $\sim 0$ \\
			\hline
			$\Lambda_{b}$ twist-$3^{+-}$ & $\sim 0$ & $-rx_{3}\Psi_{3}^{+-}\left(P_{1}-S_{1}-T_{3}-T_{7}\right)$ & $\sim 0$ & $\sim 0$ \\
			\hline
			$\Lambda_{b}$ twist-$3^{-+}$ & $\sim 0$ & $4rx_{3}\Psi_{3}^{-+}\left(P_{1}+S_{1}-T_{3}+T_{7}-2T_{2}\right)$ & $\sim 0$ & $\sim 0$ \\
			\hline
			$\Lambda_{b}$ twist-4 & $8x_3\Psi_{4}\left(A_{1}-V_{1}\right)$ & $\sim 0$ & $-8r^{2}\left(1-x_{2}^{\prime}\right)\Psi_{4}\left(V_{4}-A_{4}-2A_{5}\right)$ & $\sim 0$ \\
			\hline
			\hline
		\end{tabular*}
		\label{Tab_d7f}
	\end{table}
	
	\begin{figure}[htbp]
		\centering
		\includegraphics[width=1.0\textwidth]{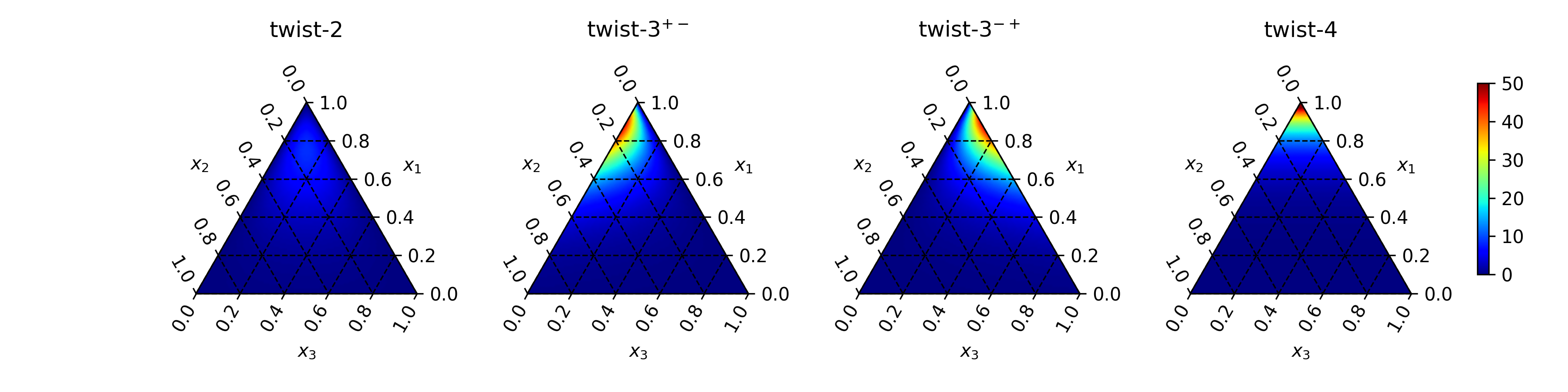}
		\captionsetup{justification=raggedright,singlelinecheck=false}
		\caption{Dependencies of the $\Lambda_b$ baryon LCDAs $\psi_2$, $\psi_3^{+-}$, $\psi_3^{-+}$ and $\psi_4$ on the momentum fractions $x_i$ in the exponential model~\cite{Bell:2013tfa}. Each point inside the triangles satisfies the relation $x_1+x_2+x_3=1$.}
		\label{lcda_lbe}
	\end{figure} 	
	
	
	To calculate the observables of semileptonic decays, extrapolation of the $\Lambda_b\to \Lambda$ form factors are performed to the whole kinematic range $0\leq q^2 \leq q_{max}^2=(m_{\Lambda_b}-m_{\Lambda})^2$. The form factors are firstly evaluated at six $q^2$ points in the low-$q^2$ region, $0\leq q^2 \leq m_{\tau}^2$, where the PQCD approach remains reliable, and then extrapolated to high-$q^2$ regime using the z-series formula~\cite{Bourrely:2008za},
	\begin{equation}
		F(q^2) = \frac{1}{1-q^2/(m_{\rm pole}^F)^2} \sum_{i=0}^k \left( a_i^F[z(q^2)]^i \right),
		\label{eq:z-series}
	\end{equation}
	where $F= f_{+,\perp,0},g_{+,\perp,0},h_{+,\perp},\widetilde{h}_{+,\perp}$, and the parameter $z$ is defined as,
	\begin{equation}
		z(q^2) = \frac{\sqrt{t_+-q^2}-\sqrt{t_+-t_0}}{\sqrt{t_+-q^2}+\sqrt{t_+-t_0}},\label{eq_z}
	\end{equation}
	with $t_0 = q^2_{\rm max} = (M_{\Lambda_b} - m_{\Lambda})^2$ and $t_+ = (m_B + m_K)^2$. The pole masses are taken from~\cite{Detmold:2016pkz} and collected in Table~\ref{table_pole}. We truncate the $z$-series expansion formula in Eq.~(\ref{eq:z-series}) at $k=1$ for simplicity,
	\begin{equation}
		F(q^2) = \frac{1}{1-q^2/(m_{\rm pole}^F)^2} \big[ a_0^F + a_1^F\:z(q^2) \big]. \label{eq:nominalfitphys}
	\end{equation}
    To constrain the behavior at high $q^2$ region, the form factors at $q^2_{max}$ from the lattice QCD calculation~\cite{Detmold:2016pkz} are included into our global fit.
	\begin{table}[htbp]
		\renewcommand{\arraystretch}{1.1} 
		\begin{tabular}{ccccc}
			\hline\hline
			$f$     & \hspace{1ex} & $J^P$ & \hspace{1ex}  & $m_{\rm pole}^f$ [GeV]  \\
			\hline
			$f_+$, $f_\perp$, $h_+$, $h_\perp$                         && $1^-$   && $5.416$  \\
			$f_0$                                                      && $0^+$   && $5.711$  \\
			$g_+$, $g_\perp$, $\widetilde{h}_+$, $\widetilde{h}_\perp$ && $1^+$   && $5.750$  \\
			$g_0$                                                      && $0^-$   && $5.367$  \\
			\hline\hline
		\end{tabular}
		\caption{Values of the pole masses, $m_{\rm pole}^f$~\cite{Detmold:2016pkz}.}
		\label{table_pole}
	\end{table}

    The central values and uncertainties of the fit parameters $a_0^F$ and $a_1^F$ are listed in Table~\ref{tab:nominal}. The corresponding form factors at the full kinematic region are depicted in Fig.~\ref{entir}.
	\begin{table}[htbp]
		\renewcommand{\arraystretch}{1.15} 
		\begin{tabular}{ccccc}
			\hline \hline
			Parameter         &  Value  & \hspace{2ex} &  Parameter         & Value  \\
			\hline
			$a_0^{f_+}$ & $\wm 0.418\pm 0.031$ &&  $a_0^{g_\perp}$ & $0.357\pm 0.027$ \\  
			$a_1^{f_+}$ & $-1.21\pm 0.15$ &&  $a_1^{g_\perp}$ & $-0.99\pm 0.14$ \\  
			$a_0^{f_0}$ & $\wm 0.370\pm 0.032$ &&  $a_0^{h_+}$ & $\wm 0.49\pm 0.05$ \\  
			$a_1^{f_0}$ & $-1.00\pm 0.16$ &&  $a_1^{h_+}$ & $-1.37\pm 0.23$ \\  
			$a_0^{f_\perp}$ & $\wm 0.52\pm 0.04$ &&  $a_0^{h_\perp}$ & $\wm 0.386\pm 0.034$ \\  
			$a_1^{f_\perp}$ & $-1.50\pm 0.19$ &&  $a_1^{h_\perp}$ & $-1.04\pm 0.17$ \\  
			$a_0^{g_+}$ & $\wm 0.356\pm 0.023$ &&  $a_0^{\widetilde{h}_+}$ & $\wm 0.338\pm 0.029$ \\  
			$a_1^{g_+}$ & $-0.89\pm 0.13$ &&  $a_1^{\widetilde{h}_+}$ & $-0.91\pm 0.15$ \\  
			$a_0^{g_0}$ & $\wm 0.397\pm 0.029$ &&  $a_0^{\widetilde{h}_\perp}$ & $0.342\pm 0.029$ \\  
			$a_1^{g_0}$ & $\wm -1.07\pm 0.15$ &&  $a_1^{\widetilde{h}_\perp}$ & $-0.85\pm 0.15$ \\  
			\hline\hline
		\end{tabular}
		\caption{Central values and uncertainties of the fit parameters.}
		\label{tab:nominal}
	\end{table}

	\begin{figure}[htbp]
		\centering
		\begin{minipage}{0.48\textwidth}
			\centering
			\includegraphics[width=\textwidth]{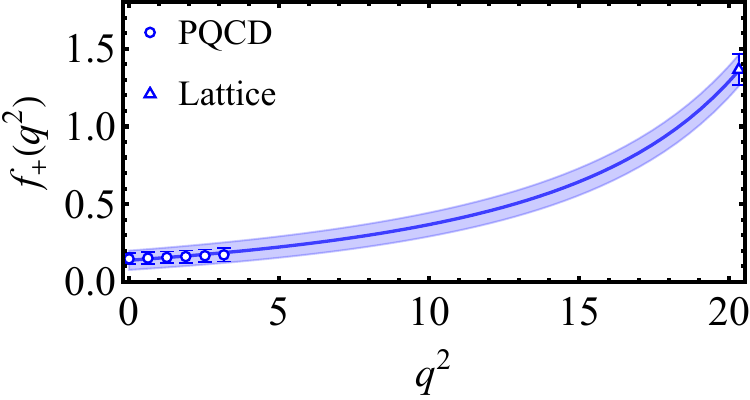}
			\label{fig:imagea1-1}
		\end{minipage}
		\hspace{0.02\textwidth} 
		\begin{minipage}{0.48\textwidth}
			\centering
			\includegraphics[width=\textwidth]{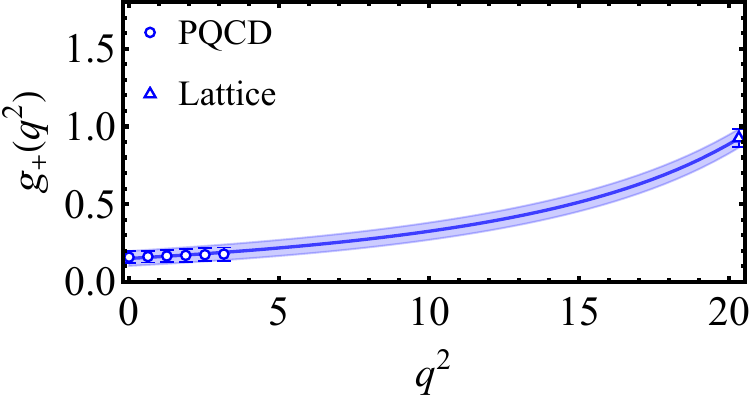}
			\label{fig:imageb1-2}
		\end{minipage}
		
		\vspace{0pt}
		
		\begin{minipage}{0.48\textwidth}
			\centering
			\includegraphics[width=\textwidth]{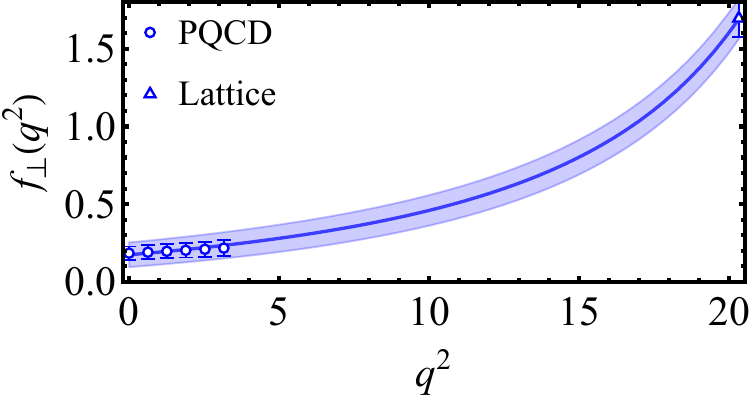}
			\label{fig:imagea2-1}
		\end{minipage}
		\hspace{0.02\textwidth}
		\begin{minipage}{0.48\textwidth}
			\centering
			\includegraphics[width=\textwidth]{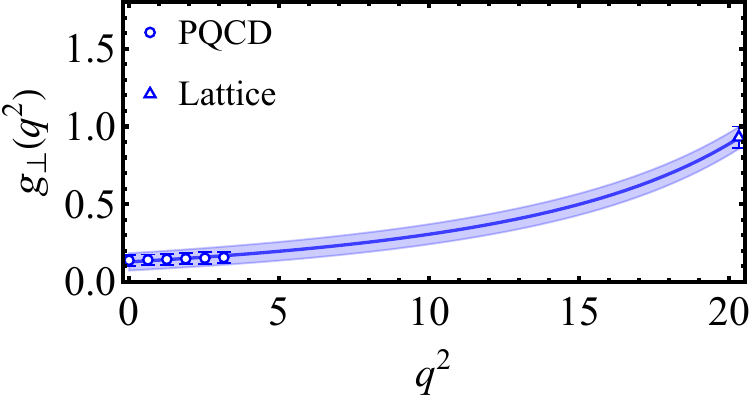}
			\label{fig:imageb2-2}
		\end{minipage}
		
		\vspace{0pt}
		
		\begin{minipage}{0.48\textwidth}
			\centering
			\includegraphics[width=\textwidth]{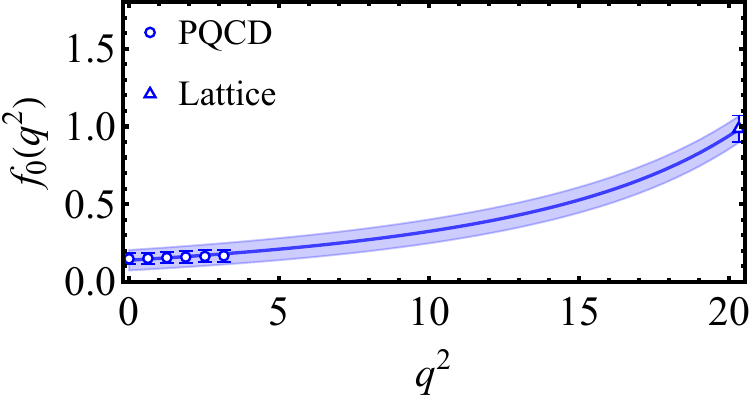}
			\label{fig:imagea3-1}
		\end{minipage}
		\hspace{0.02\textwidth}
		\begin{minipage}{0.48\textwidth}
			\centering
			\includegraphics[width=\textwidth]{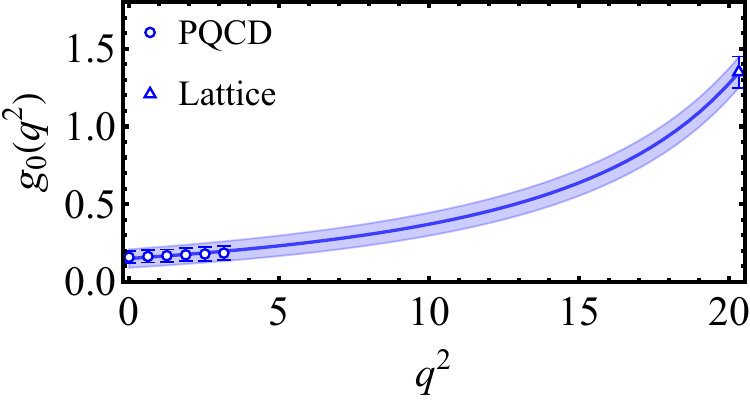}
			\label{fig:imageb3-2}
		\end{minipage}
		
		\vspace{0pt}
		
		\begin{minipage}{0.48\textwidth}
			\centering
			\includegraphics[width=\textwidth]{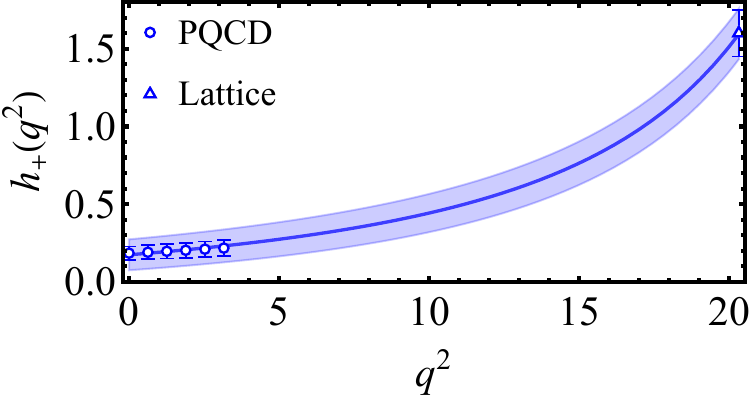}
			\label{fig:imagea4-1}
		\end{minipage}
		\hspace{0.02\textwidth}
		\begin{minipage}{0.48\textwidth}
			\centering
			\includegraphics[width=\textwidth]{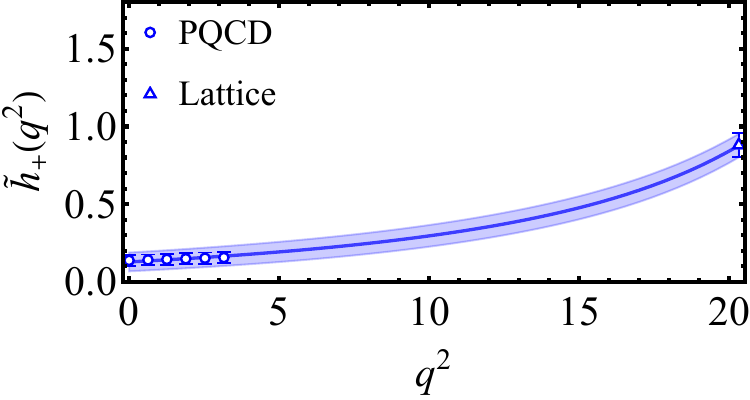}
			\label{fig:imageb4-2}
		\end{minipage}
		
		\vspace{0pt}
		
		\begin{minipage}{0.48\textwidth}
			\centering
			\includegraphics[width=\textwidth]{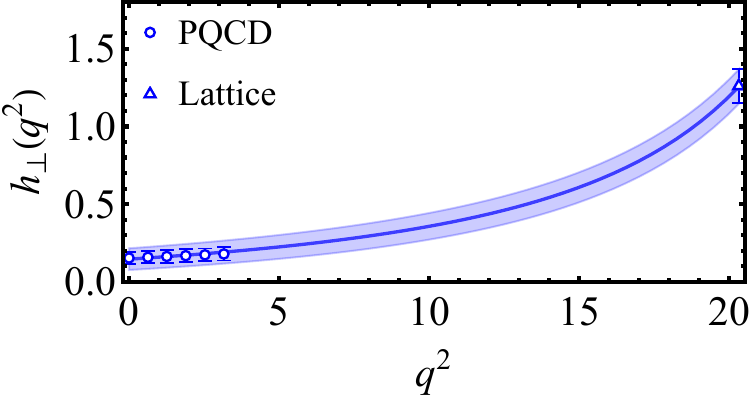}
			\label{fig:imagea5-1}
		\end{minipage}
		\hspace{0.02\textwidth}
		\begin{minipage}{0.48\textwidth}
			\centering
			\includegraphics[width=\textwidth]{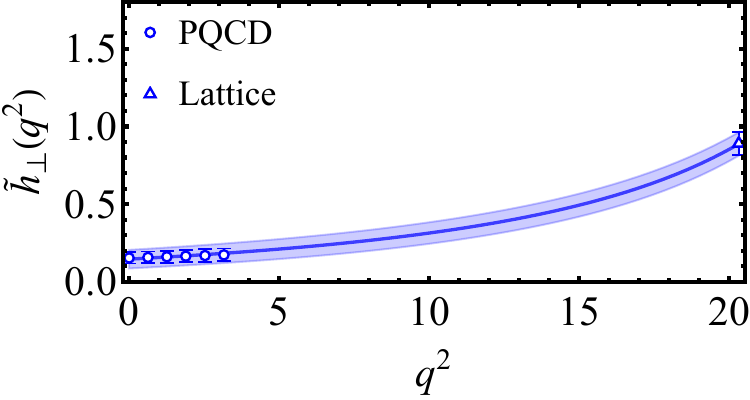}
			\label{fig:imageb5-2}
		\end{minipage}
		\captionsetup{skip=0.5pt}
		\caption{Form factors at the full kinematic range.}
		\label{entir}
	\end{figure}

\section{observables}\label{sec:observables}
	The differential decay rate of the $\Lambda_b$ decay four-body $\Lambda_b \to \Lambda(\to p^+ \pi^-) \, \ell^+ \ell^-$ with the $\Lambda_b$ unpolarized is derived as~\cite{Gutsche:2013pp,Boer:2014kda}
	\begin{eqnarray}
		\nonumber \frac{\mathrm{d}^4 \Gamma}{\mathrm{d} q^2\: \mathrm{d} \cos\theta_\ell\: \mathrm{d} \cos\theta_\Lambda\: \mathrm{d} \phi}
		&=&          \frac{3}{8\pi} \bigg[  \hspace{2.8ex}  \big( K_{1ss} \sin^2\theta_\ell +\, K_{1cc} \cos^2\theta_\ell + K_{1c} \cos\theta_\ell\big)       \\
		\nonumber && \phantom{\frac{3}{8\pi} \bigg[} + \big( K_{2ss} \sin^2\theta_\ell +\, K_{2cc} \cos^2\theta_\ell + K_{2c} \cos\theta_\ell\big) \cos\theta_\Lambda  \\
		\nonumber && \phantom{\frac{3}{8\pi} \bigg[} + \big( K_{3sc}\sin\theta_\ell \cos\theta_\ell + K_{3s} \sin\theta_\ell\big) \sin\theta_\Lambda \sin\phi          \\
		&& \phantom{\frac{3}{8\pi} \bigg[} + \big( K_{4sc}\sin\theta_\ell \cos\theta_\ell + K_{4s} \sin\theta_\ell\big) \sin\theta_\Lambda \cos\phi \bigg], \label{eq:dGamma4f}
	\end{eqnarray}
	where the angles $\theta_\Lambda$ and $\theta_\ell$ describe the polarization directions of the proton and negatively charged lepton, respectively. $K_{n\lambda}$ with $n=1,\dots,4$ and $\lambda=s,ss,c,cc,sc$ are functions of $q^2$, and its explicit expressions can be found in Eqs.(3.29)-(3.32) in Ref.~\cite{Boer:2014kda}.
	$\phi$ is the azimuthal angle between the decay planes of $p\pi$ and $\ell^+\ell^-$. By integrating over the three angles appearing in Eq.~\eqref{eq:dGamma4f}, we can obtain the $q^2$-differential decay rate as,
	\begin{equation}
		\frac{\mathrm{d}\Gamma}{\mathrm{d}q^2}=2 K_{1ss} + K_{1cc}.
	\end{equation}
	In addition to the differential decay rate, we also define the normalized angular observables,
	\begin{equation}
		\hat{K}_i = \frac{K_i}{\mathrm{d}\Gamma/\mathrm{d}q^2}.
	\end{equation}
	Then the fraction of longitudinally polarized dileptons, the lepton-side forward-backward asymmetry, the hadron-side forward-backward asymmetry, and a combined
	lepton-hadron forward-backward asymmetry can be defined as
	\begin{eqnarray}
		F_L&=&2\hat{K}_{1ss}-\hat{K}_{1cc}, \\
		A_{\rm FB}^\ell&=&\frac32\hat{K}_{1c},  \\
		A_{\rm FB}^\Lambda&=&\hat{K}_{2ss} + \frac12\hat{K}_{2cc}, \label{afbl} \\
		A_{\rm FB}^{\ell\Lambda}&=&\frac34\hat{K}_{2c}\label{afbll}.
	\end{eqnarray}
	The parameters $\hat{K}_{2ss}$, $\hat{K}_{2cc}$, and $\hat{K}_{2c}$ in Eqs.~\eqref{afbl} and~\eqref{afbll} are related to the asymmetry parameters of $\Lambda \to p\pi^-$. Here, we adopt the values provided by the PDG~\cite{ParticleDataGroup:2024cfk}, $\alpha_{-} = 0.747 \pm 0.009$.
	
	Differential branching ratio of the $\Lambda_b \to \Lambda \mu^+\mu^-$ decay obtained in this work is shown in Fig.~\ref{fig:br}. Fig.~\ref{fig_ob} presents the angular observables in the quasi four-body decay $\Lambda_b \to \Lambda(\to p \pi^-)\mu^+\mu^-$. In the current analysis, we focus only on the dominant theoretical uncertainties arising from the $\Lambda_b\to\Lambda$ form factors, the other potential sources of uncertainty are neglected as they are expected to be subdominant.
	By integrating the differential branching fraction over the kinematic range $(2m_{\mu})^2\leq q^2 \leq (M_{\Lambda_{b}}-m_{\Lambda})^2$, we obtain the total branching fraction for the $\Lambda_b\to\Lambda\mu^+\mu^-$ decay,
		\begin{eqnarray}
		\int_{(2m_{\mu})^2}^{(M_{\Lambda_{b}}-m_{\Lambda})^2}\frac{\mathrm{d}\mathcal{B}}{\mathrm{d}q^2}\mathrm{d}q^2=
			\left(1.08^{+0.15}_{-0.14}\right)\times10^{-6}.
	\end{eqnarray}
    The above prediction is in good agreement with the experimental measurement, $(1.08\pm0.28)\times10^{-6}$~\cite{ParticleDataGroup:2024cfk}.
	
	As shown in Fig.~\ref{fig:br}, the theoretical uncertainty for the differential branching ratio stemming from the form factors is large. Future improvements in determining the LCDAs of both $\Lambda_b$ and $\Lambda$ baryons could significantly reduce uncertainty of the branching ratio. The angular observables presented in Fig.~\ref{fig_ob} exhibit less sensitivity to uncertainties of the form factors due to partial cancellation in their definitions.
    \begin{figure}[htbp]
    \centering
    \includegraphics[width=0.6\textwidth]{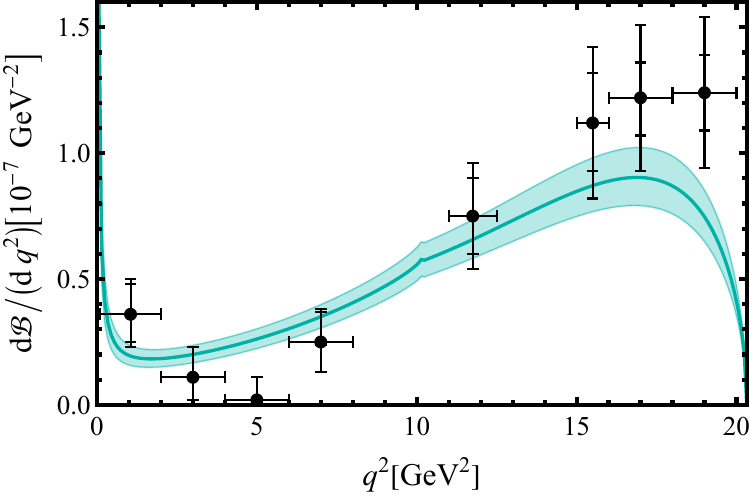}
    \captionsetup{justification=raggedright,singlelinecheck=false}
    \caption{Differential branching fraction of the $\Lambda_b \to \Lambda \mu^+\mu^-$ decay. The cyan curve and the lighter band correspond to the central values and theoretical uncertainties obtained in this work, respectively, compared with the experimental data from LHCb~\cite{LHCb:2015tgy}.}
    \label{fig:br}
    \end{figure} 
			
	\begin{figure}[htbp]
		\centering
		\begin{minipage}{0.48\textwidth}
			\centering
			\includegraphics[width=\textwidth]{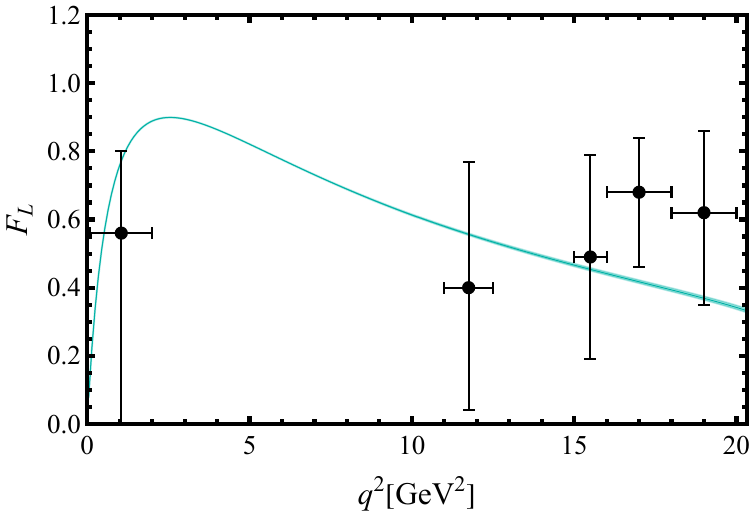}
		\end{minipage}
		\hspace{0.02\textwidth} 
		\begin{minipage}{0.48\textwidth}
			\centering
			\includegraphics[width=\textwidth]{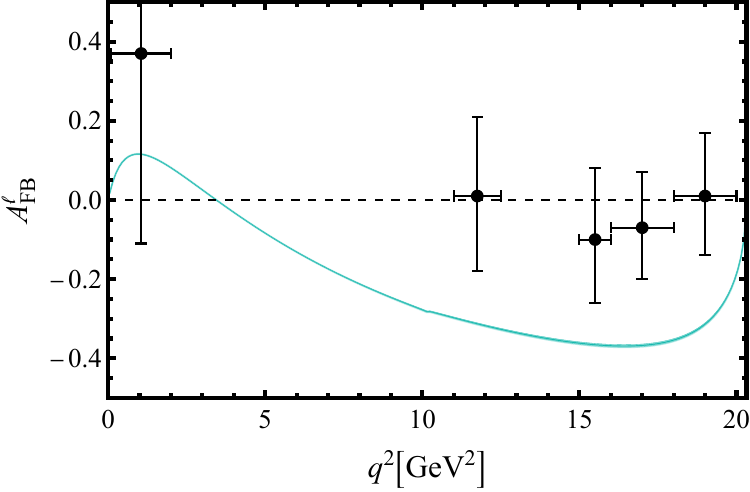}
		\end{minipage}
		
		\vspace{0pt}
		
		\begin{minipage}{0.48\textwidth}
			\centering
			\includegraphics[width=\textwidth]{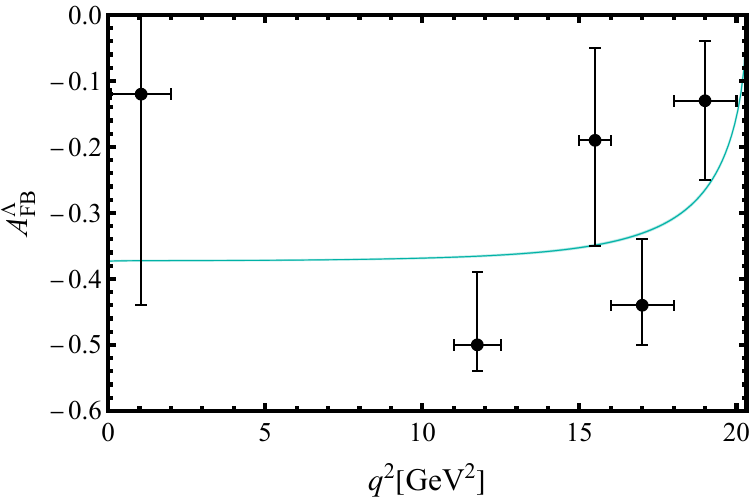}
		\end{minipage}
		\hspace{0.02\textwidth}
		\begin{minipage}{0.48\textwidth}
			\centering
			\includegraphics[width=\textwidth]{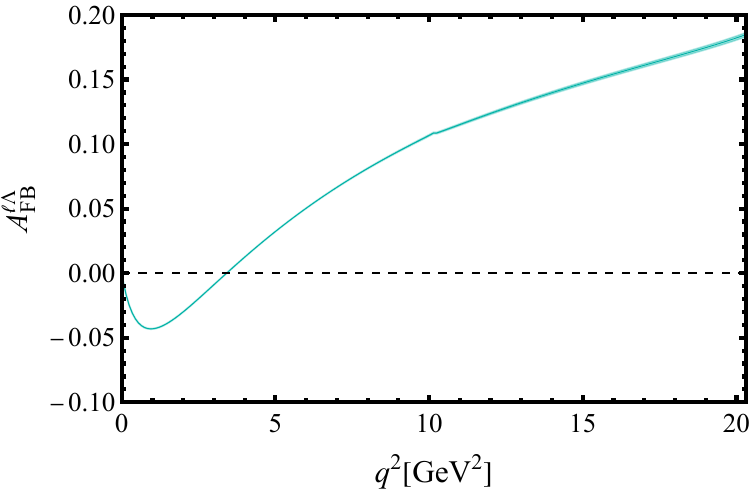}
		\end{minipage}
		\captionsetup{justification=raggedright,singlelinecheck=false}
		\caption{Angular observables of the $\Lambda_b \to \Lambda(\to p^+ \pi^-)\mu^+\mu^-$ decay obtained in this work, where the bands represent the corresponding theoretical uncertainties, compared with the experimental data from LHCb~\cite{LHCb:2015tgy}.}
		\label{fig_ob}
	\end{figure}	

		\section{Summary}\label{sec:summary}
		In this paper, we calculate the $\Lambda_b\to\Lambda$ transition form factors in the PQCD approach. The obtained form factors are close to those from other theoretical approaches within uncertainties. In our calculation, we consider the contributions of the $\Lambda_b$ LCDAs up to twist-4 and $\Lambda$ LCDAs up to twist-6, and find that high-power distribution amplitudes provide the dominant contributions to the form factors, indicating that the high-power corrections should be properly taken into account in the studies of baryon decays. In addition, we obtained the form factors at the full kinematic region by performing a z-series extrapolation, in which we combine the PQCD predictions at the low-$q^2$ region and the lattice QCD results at the $q^2_{max}$. We then evaluate the differential branching fraction of the $\Lambda_b\to\Lambda\mu^+\mu^-$ decay and angular observables of the quasi four-body $\Lambda_b \to \Lambda(\to p \pi^-)\mu^+\mu^-$ decay. Our results are consistent with the most theoretical predictions and the experimental measurements within uncertainties.
		
		\section*{Acknowledgment}
		The authors would like to thank Dr.~Dong-Hao Li for helpful discussions and valuable suggestions. 
		The work is supported by the National Natural Science Foundation of China (Grant No.~12275067,~12335003 and 12505113), 
		Science and Technology R\&D Program Joint Fund Project of Henan Province (Grant No.~225200810030),
		Science and Technology Innovation Leading Talent Support Program of Henan Province,
		National Key R\&D Program of China (Grant No.~2023YFA1606000), and the Fundamental Research Funds for the Central Universities (Grant No.~lzujbky-2023-stlt01, lzujbky-2024-oy02 and lzujbky-2025-eyt01).
		
		\appendix
		{\centering\section*{Appendix: Factorization formulas}}\label{appendix}
		We list below the factorization formulas for the form factor $f_1(q^2=0)$ from the Feynman diagrams $D_i$, $i=1$-14, Those for the other five form factors can be derived in a similar way. The last two diagrams with three-gluon vertices do not contribute due to the vanishing color factors:
		{\footnotesize
			\begin{align}
				f_1^{D_1}(q^2=0)=&\mathcal{C}M_{\Lambda_b}^3 \frac{f_{\Lambda_b}}{8N_c} \frac{1}{8\sqrt{2}N_c}\int[dx]\int[dx^\prime]16\pi^2\alpha_s^2(t^{D_1})h_{f_{1}} \frac{1}{(2\pi)^5}\int b_1^\prime db_1^\prime \int b_2 db_2\nonumber\\
				&\int b_3 db_3\int d\theta_1\int d\theta_2\exp[-S^{D_1}(x,x^\prime ,b,b^\prime)]F_1(D,b_3)F_3(A,B,C,b_1^\prime,b_2),\\
				f_1^{D_2}(q^2=0)=&\mathcal{C}M_{\Lambda_b}^3 \frac{f_{\Lambda_b}}{8N_c} \frac{1}{8\sqrt{2}N_c}\int[dx]\int[dx^\prime]16\pi^2\alpha_s^2(t^{D_2})h_{f_{1}} \frac{1}{(2\pi)^5}\int b_1^\prime db_1^\prime \int b_2 db_2\nonumber\\
				&\int b_3 db_3\int d\theta_1\int d\theta_2\exp[-S^{D_2}(x,x^\prime ,b,b^\prime)]F_1(C,b_2)F_3(A,B,D,b_1^\prime,b_3),\\
				f_1^{D_3}(q^2=0)=&\mathcal{C}M_{\Lambda_b}^3 \frac{f_{\Lambda_b}}{8N_c} \frac{1}{8\sqrt{2}N_c}\int[dx]\int[dx^\prime]16\pi^2\alpha_s^2(t^{D_3})h_{f_{1}} \frac{1}{(2\pi)^7}\int b_1db_1\int b_1^\prime db_1^\prime \int b_3 db_3\nonumber\\
				&\int b_3^\prime db_3^\prime\int d\theta_1\int d\theta_2\int d\theta_3\exp[-S^{D_3}(x,x^\prime ,b,b^\prime)]F_1\left(A,\left|\vec{b_1}+\vec{b_1^{\prime}}-\vec{b_3}-\vec{b_3^\prime}\right|\right)\nonumber\\
				&F_1\left(B,\left|\vec{b_3}+\vec{b_3^\prime}\right|\right)F_1(C,b_1)F_1(D,b_3^\prime),\\
				f_1^{D_4}(q^2=0)=&\mathcal{C}M_{\Lambda_b}^3 \frac{f_{\Lambda_b}}{8N_c} \frac{1}{8\sqrt{2}N_c}\int[dx]\int[dx^\prime]16\pi^2\alpha_s^2(t^{D_4})h_{f_{1}} \frac{1}{(2\pi)^7}\int b_1db_1\int b_1^\prime db_1^\prime \int b_3 db_3\nonumber\\
				&\int b_3^\prime db_3^\prime\int d\theta_1\int d\theta_2\int d\theta_3\exp[-S^{D_4}(x,x^\prime ,b,b^\prime)]F_1\left(A,\left|\vec{b_3}+\vec{b_3^{\prime}}-\vec{b_1}-\vec{b_1^\prime}\right|\right)\nonumber\\
				&F_1\left(B,\left|\vec{b_3}+\vec{b_3^\prime}\right|\right)F_1\left(C,\left|\vec{b_3}+\vec{b_3^\prime}-\vec{b_1}\right|\right)F_1(D,b_3),\\
				f_1^{D_5}(q^2=0)=&\mathcal{C}M_{\Lambda_b}^3 \frac{f_{\Lambda_b}}{8N_c} \frac{1}{8\sqrt{2}N_c}\int[dx]\int[dx^\prime]16\pi^2\alpha_s^2(t^{D_5})h_{f_{1}} \frac{1}{(2\pi)^7}\int b_1db_1\int b_1^\prime db_1^\prime \int b_2 db_2\nonumber\\
				&\int b_2^\prime db_2^\prime\int d\theta_1\int d\theta_2\int d\theta_3\exp[-S^{D_5}(x,x^\prime ,b,b^\prime)]F_1\left(A,\left|\vec{b_2}+\vec{b_2^\prime}-\vec{b_1}-\vec{b_1^\prime}\right|\right)\nonumber\\
				&F_1\left(B,\left|\vec{b_2}+\vec{b_2^\prime}\right|\right)F_1\left(C,\left|\vec{b_2}+\vec{b_2^\prime}-\vec{b_1}\right|\right)F_1(D,b_2),\\
				f_1^{D_6}(q^2=0)=&\mathcal{C}M_{\Lambda_b}^3 \frac{f_{\Lambda_b}}{8N_c} \frac{1}{8\sqrt{2}N_c}\int[dx]\int[dx^\prime]16\pi^2\alpha_s^2(t^{D_6})h_{f_{1}} \frac{1}{(2\pi)^7}\int b_1db_1\int b_1^\prime db_1^\prime \int b_2 db_2\nonumber\\
				&\int b_2^\prime db_2^\prime\int d\theta_1\int d\theta_2\int d\theta_3\exp[-S^{D_6}(x,x^\prime ,b,b^\prime)]F_1\left(A,\left|\vec{b_2}+\vec{b_2^\prime}-\vec{b_1}-\vec{b_1^\prime}\right|\right)\nonumber\\
				&F_1\left(B,\left|\vec{b_2}+\vec{b_2^\prime}\right|\right)F_1(C,b_2^\prime)F_1(D,b_1),\\
				f_1^{D_7}(q^2=0)=&\mathcal{C}M_{\Lambda_b}^3 \frac{f_{\Lambda_b}}{8N_c} \frac{1}{8\sqrt{2}N_c}\int[dx]\int[dx^\prime]16\pi^2\alpha_s^2(t^{D_7})h_{f_{1}} \frac{1}{(2\pi)^5}\int b_2 db_2 \int b_3 db_3\nonumber\\
				&\int b_3^{\prime} db_3^{\prime}\int d\theta_1\int d\theta_2\exp[-S^{D_7}(x,x^\prime ,b,b^\prime)]F_1(D,b_3^\prime)F_3\left(A,C,B,\left|\vec{b_{3}}+\vec{b_3^\prime}\right|,b_2\right),\\
				f_1^{D_8}(q^2=0)=&\mathcal{C}M_{\Lambda_b}^3 \frac{f_{\Lambda_b}}{8N_c} \frac{1}{8\sqrt{2}N_c}\int[dx]\int[dx^\prime]16\pi^2\alpha_s^2(t^{D_8})h_{f_{1}} \frac{1}{(2\pi)^5}\int b_2 db_2 \int b_2^{\prime} db_2^{\prime}\nonumber\\
				&\int b_3 db_3 \int d\theta_1\int d\theta_2\exp[-S^{D_8}(x,x^\prime ,b,b^\prime)]F_1(C,b_2^\prime)F_3\left(B,D,A,\left|\vec{b_2}+\vec{b_2^\prime}\right|,b_3\right),\\
				f_1^{D_9}(q^2=0)=&\mathcal{C}M_{\Lambda_b}^3 \frac{f_{\Lambda_b}}{8N_c} \frac{1}{8\sqrt{2}N_c}\int[dx]\int[dx^\prime]16\pi^2\alpha_s^2(t^{D_9})h_{f_{1}} \frac{1}{(2\pi)^7}\int b_1db_1\int b_1^\prime db_1^\prime \int b_3 db_3\nonumber\\
				&\int b_3^\prime db_3^\prime\int d\theta_1\int d\theta_2\int d\theta_3\exp[-S^{D_9}(x,x^\prime ,b,b^\prime)]F_1\left(A,\left|\vec{b_3}+\vec{b_3^\prime}-\vec{b_1}-\vec{b_1^\prime}\right|\right)\nonumber\\
				&F_1\left(B,\left|\vec{b_3}+\vec{b_3^\prime}\right|\right)F_1\left(C,\left|\vec{b_3}+\vec{b_3^\prime}-\vec{b_1^\prime}\right|\right)F_1(D,b_3^\prime),\\
				f_1^{D_{10}}(q^2=0)=&\mathcal{C}M_{\Lambda_b}^3 \frac{f_{\Lambda_b}}{8N_c} \frac{1}{8\sqrt{2}N_c}\int[dx]\int[dx^\prime]16\pi^2\alpha_s^2(t^{D_{10}})h_{f_{1}}\frac{1}{(2\pi)^7}\int b_1db_1\int b_1^\prime db_1^\prime \int b_3 db_3\nonumber\\
				&\int b_3^\prime db_3^\prime\int d\theta_1\int d\theta_2\int d\theta_3\exp[-S^{D_{10}}(x,x^\prime ,b,b^\prime)]F_1\left(A,\left|\vec{b_3}+\vec{b_3^\prime}-\vec{b_1}-\vec{b_1^\prime}\right|\right)\nonumber\\
				&F_1\left(B,\left|\vec{b_3}+\vec{b_3^\prime}\right|\right)F_1(C,b_1^\prime)F_1(D,b_3),\\
				f_1^{D_{11}}(q^2=0)=&\mathcal{C}M_{\Lambda_b}^3 \frac{f_{\Lambda_b}}{8N_c} \frac{1}{8\sqrt{2}N_c}\int[dx]\int[dx^\prime]16\pi^2\alpha_s^2(t^{D_{11}})h_{f_{1}} \frac{1}{(2\pi)^7}\int b_1db_1\int b_1^\prime db_1^\prime \int b_2 db_2\nonumber\\
				&\int b_2^\prime db_2^\prime\int d\theta_1\int d\theta_2\int d\theta_3\exp[-S^{D_{11}}(x,x^\prime ,b,b^\prime)]F_1\left(A,\left|\vec{b_2}+\vec{b_2^\prime}-\vec{b_1}-\vec{b_1^\prime}\right|\right)\nonumber\\
				&F_1\left(B,\left|\vec{b_2}+\vec{b_2^\prime}\right|\right)F_1(C,b_2^\prime)F_1\left(D,\left|\vec{b_2}+\vec{b_2^\prime}-\vec{b_1^\prime}\right|\right),\\
				f_1^{D_{12}}(q^2=0)=&\mathcal{C}M_{\Lambda_b}^3 \frac{f_{\Lambda_b}}{8N_c} \frac{1}{8\sqrt{2}N_c}\int[dx]\int[dx^\prime]16\pi^2\alpha_s^2(t^{D_{12}})h_{f_{1}} \frac{1}{(2\pi)^7}\int b_1db_1\int b_1^\prime db_1^\prime \int b_2 db_2\nonumber\\
				&\int b_2^\prime db_2^\prime\int d\theta_1\int d\theta_2\int d\theta_3\exp[-S^{D_{12}}(x,x^\prime ,b,b^\prime)]F_1\left(A,\left|\vec{b_2}+\vec{b_2^\prime}-\vec{b_1}-\vec{b_1^\prime}\right|\right)\nonumber\\
				&F_1\left(B,\left|\vec{b_2}+\vec{b_2^\prime}\right|\right)F_1(C,b_2)F_1(D,b_1^{\prime}),\\
				f_1^{D_{13}}(q^2=0)=&\mathcal{C}M_{\Lambda_b}^3 \frac{f_{\Lambda_b}}{8N_c} \frac{1}{8\sqrt{2}N_c}\int[dx]\int[dx^\prime]16\pi^2\alpha_s^2(t^{D_{13}})h_{f_{1}} \frac{1}{(2\pi)^5}\int b_2 db_2 \int b_2^\prime db_2^\prime\nonumber\\
				&\int b_3 db_3\int d\theta_1\int d\theta_2\exp[-S^{D_{13}}(x,x^\prime ,b,b^\prime)]F_2\left(A,B,\left|\vec{b_2}+\vec{b_2^\prime}\right|\right)F_1(C,b_2^\prime)F_1(D,b_3),\\
				f_1^{D_{14}}(q^2=0)=&\mathcal{C}M_{\Lambda_b}^3 \frac{f_{\Lambda_b}}{8N_c} \frac{1}{8\sqrt{2}N_c}\int[dx]\int[dx^\prime]16\pi^2\alpha_s^2(t^{D_{14}})h_{f_{1}} \frac{1}{(2\pi)^5}\int b_2 db_2 \int b_2^\prime db_2^\prime\nonumber\\
				&\int b_3 db_3\int d\theta_1\int d\theta_2\exp[-S^{D_{14}}(x,x^\prime ,b,b^\prime)]F_2\left(A,B,\left|\vec{b_2}+\vec{b_2^\prime}\right|\right)F_1(C,b_2)F_1(D,b_3^\prime),	
		\end{align}}
		where $\mathcal{C}=8/3$ is the color factor and $[dx]\equiv dx_1dx_2dx_3\delta(1-x_1-x_2-x_3)$, $[dx']$ is defined analogously, the auxiliary functions $A,B,C$ and $D$ in Table~\ref{table:ABCD} are related to the denominators of the four propagators in each diagram. The exponent $S^{D_i}$ is the sum of the total exponents from the $\Lambda_b$ and $\Lambda$ wave functions with the hard scale $t^{D_i}$ involved in the diagram $D_i$. The functions $F_1, F_2$ and $F_3$ are written, in terms of the Fourier integrals, as
		{\footnotesize
			\begin{align}
				F_1(A,b)=&\int d^2k_T\frac{e^{i\textbf{k}_T\cdot\textbf{b}}}{k^2+A}=2\pi \left\{K_0(\sqrt{A}b)\theta(A)+\frac{\pi i}{2}\left[J_0(\sqrt{-A}b)+iN_0(\sqrt{-A}b)\right]\theta(-A)\right\},\\
				F_2(A,B,b)=&\int d^2k_T\frac{e^{i\textbf{k}_T\cdot\textbf{b}}}{(k^2+A)(k^2+B)}\nonumber\\
				=&\pi\int_{0}^{1}dz\frac{b}{\sqrt{|Z_1|}}\left\{K_1(\sqrt{Z_1}b)\theta(Z_1)
				+\frac{\pi}{2}\left[N_1(\sqrt{-Z_1}b)-iJ_1(\sqrt{-Z_1}b)\right]\theta(-Z_1)\right\},\\
				F_3(A,B,C,b_1,b_2)=&\int d^2k_{1T}\int d^2k_{2T}\frac{e^{i(\textbf{k}_{1T}\cdot\textbf{b}_1+\textbf{k}_{2T}\cdot\textbf{b}_2)}}{(k_1^2+A)(k_2^2+B)((k_1+k_2)^2+C)}\nonumber\\
				=&\pi^2\int_{0}^{1}\frac{dz_1dz_2}{z_1(1-z_1)}\frac{\sqrt{X_2}}{\sqrt{|Z_2|}}\nonumber\\
				&\times\left\{K_1(\sqrt{X_2Z_2})\theta(Z_2)+\frac{\pi}{2}\left[N_1(\sqrt{-X_2Z_2})-iJ_1(\sqrt{-X_2Z_2})\right]\theta(-Z_2)\right\},
		\end{align}}
		in which $J_n$ ($N_n$) is the Bessel function of the first (second) kind, $K_n$ follows the relation
		\begin{align}
			K_n(-iz)=\frac{\pi i}{2}e^{(in\pi)/2}\left[J_n(z)+iN_n(z)\right],
		\end{align}
		and the variables $Z_1$, $Z_2$ and $X_2$ are given by
		\begin{align}
			Z_1=&Az+B(1-z),\\
			Z_2=&A(1-z_2)+\frac{z_2}{z_1(1-z_1)}\left[B(1-z_1)+Cz_1\right],\\
			X_2=&(b_1-z_1b_2)^2+\frac{z_1(1-z_1)}{z_2}b_2^2,
		\end{align}
		with Feynman parameters $z$'s. 
		Because $A$, $Z_1$, $Z_2$ and $X_2$ are all positive, the imaginary parts of the above functions $F_{1,2,3}$ do not contribute.
		\begin{table}
			\centering
			\renewcommand\arraystretch{1.1}
			\caption{Auxiliary functions $A,B,C$ and $D$. An overall coefficient $m_{\Lambda_b}^2$ is implicit for the entries.}
			\begin{tabular*}{158mm}{c@{\extracolsep{\fill}}cccc}
				\hline
				\hline
				&A&B&C&D\\
				\hline
				$D_1$&$1-x_1^\prime$&$x_{2}+x_{3}^{\prime}-x_{2}x_{3}^{\prime}$&$x_2x_2^\prime$&$x_3x_3^\prime$\\
				$D_2$&$1-x_1^\prime$&$x_{2}^{\prime}+x_{3}-x_{2}^{\prime}x_{3}$&$x_2x_2^\prime$&$x_3x_3^\prime$\\
				$D_3$&$1-x_1^\prime$&$\left(1-x_{1}^{\prime}\right)x_{3}$&$\left(1-x_{1}\right)\left(1-x_{1}^{\prime}\right)$&$x_3x_3^\prime$\\
				$D_4$&$1-x_1^\prime$&$x_3^\prime\left(1-x_{1}\right)$&$\left(1-x_{1}\right)\left(1-x_{1}^{\prime}\right)$&$x_3x_3^\prime$\\
				$D_5$&$1-x_1^\prime$&$x_2^\prime\left(1-x_{1}\right)$&$\left(1-x_{1}\right)\left(1-x_{1}^{\prime}\right)$&$x_2x_2^\prime$\\
				$D_6$&$1-x_1^\prime$&$x_2\left(1-x_{1}^{\prime}\right)$&$x_2x_2^\prime$&$\left(1-x_{1}\right)\left(1-x_{1}^{\prime}\right)$\\
				$D_7$&$x_3\left(1-x_{2}^{\prime}\right)$&$1-x_1$&$x_2x_2^\prime$&$x_3x_3^\prime$\\
				$D_8$&$x_2\left(1-x_{3}^{\prime}\right)$&$1-x_1$&$x_2x_2^\prime$&$x_3x_3^\prime$\\
				$D_9$&$1-x_1$&$x_3\left(1-x_{1}^{\prime}\right)$&$\left(1-x_{1}\right)\left(1-x_{1}^{\prime}\right)$&$x_3x_3^\prime$\\
				$D_{10}$&$1-x_1$&$x_3^\prime\left(1-x_{1}\right)$&$\left(1-x_{1}\right)\left(1-x_{1}^{\prime}\right)$&$x_3x_3^\prime$\\
				$D_{11}$&$1-x_1$&$x_2\left(1-x_{1}^{\prime}\right)$&$x_2x_2^\prime$&$\left(1-x_{1}\right)\left(1-x_{1}^{\prime}\right)$\\
				$D_{12}$&$1-x_1$&$x_2^\prime\left(1-x_{1}\right)$&$x_2x_2^\prime$&$\left(1-x_{1}\right)\left(1-x_{1}^{\prime}\right)$\\
				$D_{13}$&$x_2\left(1-x_{3}^{\prime}\right)$&$x_{2}+x_{3}^{\prime}-x_{2}x_{3}^{\prime}$&$x_2x_2^\prime$&$x_3x_3^\prime$\\
				$D_{14}$&$x_3\left(1-x_{2}^{\prime}\right)$&$x_{2}^{\prime}+x_{3}-x_{2}^{\prime}x_{3}$&$x_2x_2^\prime$&$x_3x_3^\prime$\\
				\hline
				\hline
			\end{tabular*}\label{table:ABCD}
		\end{table}



\begin{thebibliography}{99}
	
	
	\bibitem{Glashow:1970gm}
	S.~L.~Glashow, J.~Iliopoulos and L.~Maiani,
	Phys. Rev. D \textbf{2} (1970), 1285-1292.
	
	\bibitem{LHCb:2014cxe}
	R.~Aaij \textit{et al.} [LHCb],
	JHEP \textbf{06} (2014), 133.
	
	\bibitem{LHCb:2016ykl}
	R.~Aaij \textit{et al.} [LHCb],
	JHEP \textbf{11} (2016), 047.
	
	\bibitem{LHCb:2021zwz}
	R.~Aaij \textit{et al.} [LHCb],
	Phys. Rev. Lett. \textbf{127} (2021) no.15, 151801.
	
	\bibitem{LHCb:2015tgy}
	R.~Aaij \textit{et al.} [LHCb],
	JHEP \textbf{06} (2015), 115.
	
\bibitem{LHCb:2025ray}
R.~Aaij \textit{et al.} [LHCb],
Nature \textbf{643} (2025) no.8074, 1223-1228.
	
	\bibitem{Wang:2024oyi}
	J.~P.~Wang and F.~S.~Yu,
	Chin. Phys. C \textbf{48} (2024) no.10, 101002.
	
\bibitem{Yu:2025ekh}
F.~S.~Yu and C.~D.~L{\"u},
Sci. Bull. \textbf{70} (2025), 2035-2036.
	
	\bibitem{LHCb:2024yzj}
	R.~Aaij \textit{et al.} [LHCb],
	Phys. Rev. Lett. \textbf{134} (2025) no.10, 101802.
	
	\bibitem{Feldmann:2011xf}
	T.~Feldmann and M.~W.~Y.~Yip,
	Phys. Rev. D \textbf{85} (2012), 014035.
	
	\bibitem{Detmold:2012vy}
	W.~Detmold, C.~J.~D.~Lin, S.~Meinel and M.~Wingate,
	Phys. Rev. D \textbf{87} (2013) no.7, 074502.
	
	\bibitem{Detmold:2016pkz}
	W.~Detmold and S.~Meinel,
	Phys. Rev. D \textbf{93} (2016) no.7, 074501.
	
	\bibitem{Wang:2015ndk}
	Y.~M.~Wang and Y.~L.~Shen,
	JHEP \textbf{02} (2016), 179.
	
	\bibitem{Kurimoto:2001zj}
	T.~Kurimoto, H.~n.~Li and A.~I.~Sanda,
	Phys. Rev. D \textbf{65}, 014007 (2002).
	
	\bibitem{Li:2012nk}
	H.~n.~Li, Y.~L.~Shen and Y.~M.~Wang,
	Phys. Rev. D \textbf{85}, 074004 (2012).
	
	\bibitem{Lu:2000em}
	C.~D.~Lu, K.~Ukai and M.~Z.~Yang,
	Phys. Rev. D \textbf{63}, 074009 (2001).
	
	\bibitem{Cheng:2014rfa}
	H.~Y.~Cheng, C.~W.~Chiang and A.~L.~Kuo,
	Phys. Rev. D \textbf{91}, no.1, 014011 (2015).
	
	\bibitem{Belle:2004mad}
	K.~Abe \textit{et al.} [Belle],
	Phys. Rev. Lett. \textbf{93} (2004), 021601.
	
	\bibitem{BaBar:2004gyj}
	B.~Aubert \textit{et al.} [BaBar],
	Phys. Rev. Lett. \textbf{93} (2004), 131801.
	
	\bibitem{Belle:2004nch}
	Y.~Chao \textit{et al.} [Belle],
	Phys. Rev. Lett. \textbf{93} (2004), 191802.
	
	\bibitem{LHCb:2018fly}
	R.~Aaij \textit{et al.} [LHCb],
	Phys. Lett. B \textbf{787} (2018), 124-133.
	
	\bibitem{LHCb:2024iis}
	R.~Aaij \textit{et al.} [LHCb],
	Phys. Rev. D \textbf{111} (2025) no.9, 092004.
	
	\bibitem{Han:2024kgz}
	J.~J.~Han, J.~X.~Yu, Y.~Li, H.~n.~Li, J.~P.~Wang, Z.~J.~Xiao and F.~S.~Yu,
	Phys. Rev. Lett. \textbf{134} (2025) no.22, 221801.
	
\bibitem{Han:2025tvc}
J.~J.~Han, J.~X.~Yu, Y.~Li, H.~n.~Li, J.~P.~Wang, Z.~J.~Xiao and F.~S.~Yu,
Phys. Rev. D \textbf{112} (2025) no.5, 053007.
	

	\bibitem{Li:1992ce}
	H.~n.~Li,
	Phys. Rev. D \textbf{48} (1993), 4243-4254.
	
	\bibitem{Kundu:1998gv}
	B.~Kundu, H.~n.~Li, J.~Samuelsson and P.~Jain,
	Eur. Phys. J. C \textbf{8} (1999), 637-642.
	
	\bibitem{Shih:1998pb}
	H.~H.~Shih, S.~C.~Lee and H.~n.~Li,
	Phys. Rev. D \textbf{59} (1999), 094014.
	
	\bibitem{Lu:2009cm}
	C.~D.~Lu, Y.~M.~Wang, H.~Zou, A.~Ali and G.~Kramer,
	Phys. Rev. D \textbf{80} (2009), 034011.
	
	\bibitem{Han:2022srw}
	J.~J.~Han, Y.~Li, H.~n.~Li, Y.~L.~Shen, Z.~J.~Xiao and F.~S.~Yu,
	Eur. Phys. J. C \textbf{82} (2022) no.8, 686.
	
	\bibitem{Rui:2022sdc}
	Z.~Rui, C.~Q.~Zhang, J.~M.~Li and M.~K.~Jia,
	Phys. Rev. D \textbf{106} (2022) no.5, 053005.
	
	\bibitem{Bali:2015ykx}
	G.~S.~Bali, V.~M.~Braun, M.~G{\"o}ckeler, M.~Gruber, F.~Hutzler, A.~Sch{\"a}fer, R.~W.~Schiel, J.~Simeth, W.~S{\"o}ldner and A.~Sternbeck, \textit{et al.}
	JHEP \textbf{02} (2016), 070.
	
	\bibitem{RQCD:2019hps}
	G.~S.~Bali \textit{et al.} [RQCD],
	Eur. Phys. J. A \textbf{55} (2019) no.7, 116.
	
	\bibitem{Liu:2014uha}
	Y.~L.~Liu, C.~Y.~Cui and M.~Q.~Huang,
	Eur. Phys. J. C \textbf{74} (2014), 3041.
	
	\bibitem{El-Khadra:2001wco}
	A.~X.~El-Khadra, A.~S.~Kronfeld, P.~B.~Mackenzie, S.~M.~Ryan and J.~N.~Simone,
	Phys. Rev. D \textbf{64} (2001), 014502.
	
	\bibitem{Gutsche:2013pp}
	T.~Gutsche, M.~A.~Ivanov, J.~G.~Korner, V.~E.~Lyubovitskij and P.~Santorelli,
	Phys. Rev. D \textbf{87} (2013), 074031.
	
	\bibitem{Li:2001ay}
	H.~n.~Li,
	Phys. Rev. D \textbf{66} (2002), 094010.
	
	\bibitem{Lu:2000hj}
	C.~D.~Lu and M.~Z.~Yang,
	Eur. Phys. J. C \textbf{23} (2002), 275-287.
	
	\bibitem{Bell:2013tfa}
	G.~Bell, T.~Feldmann, Y.~M.~Wang and M.~W.~Y.~Yip,
	JHEP \textbf{11} (2013), 191.
	
	\bibitem{Ball:2008fw}
	P.~Ball, V.~M.~Braun and E.~Gardi,
	Phys. Lett. B \textbf{665} (2008), 197-204.
	
	\bibitem{Ali:2012zza}
	A.~Ali, C.~Hambrock and A.~Y.~Parkhomenko,
	Theor. Math. Phys. \textbf{170} (2012), 2-16.
	
	\bibitem{Wein:2015oqa}
	P.~Wein and A.~Sch{\"a}fer,
	JHEP \textbf{05} (2015), 073.
	
	\bibitem{Braun:2000kw}
	V.~Braun, R.~J.~Fries, N.~Mahnke and E.~Stein,
	Nucl. Phys. B \textbf{589} (2000), 381-409.
	
	\bibitem{Liu:2009mb}
	Y.~L.~Liu and M.~Q.~Huang,
	Phys. Rev. D \textbf{79} (2009), 114031.
	
	\bibitem{LatticeParton:2024vck}
	M.~H.~Chu \textit{et al.} [Lattice Parton],
	Phys. Rev. D \textbf{111} (2025) no.3, 034510.
	
	\bibitem{LPC:2025jvd}
	H.~Bai \textit{et al.} [LPC],
	[arXiv:2508.08971 [hep-lat]].
	
	\bibitem{Aliev:2010uy}
	T.~M.~Aliev, K.~Azizi and M.~Savci,
	Phys. Rev. D \textbf{81} (2010), 056006.
	
	\bibitem{Faustov:2017wbh}
	R.~N.~Faustov and V.~O.~Galkin,
	Phys. Rev. D \textbf{96} (2017) no.5, 053006.
	
	\bibitem{Gutsche:2013oea}
	T.~Gutsche, M.~A.~Ivanov, J.~G.~K{\"o}rner, V.~E.~Lyubovitskij and P.~Santorelli,
	Phys. Rev. D \textbf{88} (2013) no.11, 114018.
	
	\bibitem{Huang:1998ek}
	C.~S.~Huang and H.~G.~Yan,
	Phys. Rev. D \textbf{59} (1999), 114022.
	
	\bibitem{Zhu:2018jet}
	J.~Zhu, Z.~T.~Wei and H.~W.~Ke,
	Phys. Rev. D \textbf{99} (2019) no.5, 054020.
	
	\bibitem{Mott:2011cx}
	L.~Mott and W.~Roberts,
	Int. J. Mod. Phys. A \textbf{27} (2012), 1250016.
	

	
	\bibitem{Wei:2009np}
	Z.~T.~Wei, H.~W.~Ke and X.~Q.~Li,
	Phys. Rev. D \textbf{80} (2009), 094016.
	
	
	
	\bibitem{Mohanta:2000nk}
	R.~Mohanta, A.~K.~Giri and M.~P.~Khanna,
	Phys. Rev. D \textbf{63} (2001), 074001.
	
	
	\bibitem{Cheng:1996cs}
	H.~Y.~Cheng,
	Phys. Rev. D \textbf{56} (1997), 2799-2811.
	
	\bibitem{Wang:2011uv}
	W.~Wang,
	Phys. Lett. B \textbf{708} (2012), 119-126.
	
	\bibitem{Bourrely:2008za}
	C.~Bourrely, I.~Caprini and L.~Lellouch,
	Phys. Rev. D \textbf{79} (2009), 013008.
	
	\bibitem{Boer:2014kda}
	P.~B{\"o}er, T.~Feldmann and D.~van Dyk,
	JHEP \textbf{01}, 155 (2015).


\bibitem{ParticleDataGroup:2024cfk}
S.~Navas \textit{et al.} [Particle Data Group],
Phys. Rev. D \textbf{110} (2024) no.3, 030001.
		

	
	
\end{thebibliography}
	\end{document}